\documentclass[aps,showpacs,preprint,groupedaddress,amsmath,amssymb,superscriptaddress,letterpaper]{revtex4}
\usepackage{graphicx}
\usepackage{dcolumn} 
\usepackage{bm}      
\usepackage{times}
\usepackage{color}
\usepackage{aas_macros}

\usepackage{setspace}

\newcommand{\lansa}{{LANS} }

\newcommand{\lans}{LANS}
\newcommand{\lamhd}{LAMHD}
\newcommand{\lamhda}{{LAMHD} }

\newcommand{\dof}{{$N_{dof}$}}

\renewcommand{\vec}{\mathbf}

\def\vier#1{{#1}} 
\def\neu#1{{#1}} 
\newcommand{\resp}[2]{{#1}{#2}}
\newcommand{\add}[2]{{#1}{#2}}
\def\AD#1{{#1}} 

\begin{document}

\title{ 
The Lagrangian-averaged model for
  magnetohydrodynamics turbulence and the absence of bottleneck}
 
 \author{Jonathan \surname{Pietarila Graham}}
 \affiliation{Max-Planck-Institut f\"ur 
Sonnensystemforschung, 37191 Katlenburg-Lindau, Germany}
  \author{Pablo D. Mininni}
 \affiliation{National
  Center for Atmospheric Research,\footnote{The National Center for
  Atmospheric Research is sponsored by the National Science
  Foundation} P.O. Box 3000, Boulder, Colorado 80307, USA}
 \affiliation{Departamento 
de F\'\i sica, Facultad de Ciencias Exactas y Naturales, Universidad de 
Buenos Aires, Ciudad Universitaria, 1428 Buenos Aires, Argentina}
  \author{Annick Pouquet} 
 \affiliation{National
  Center for Atmospheric Research,\footnote{The National Center for
  Atmospheric Research is sponsored by the National Science
  Foundation} P.O. Box 3000, Boulder, Colorado 80307, USA}
 
 \date{\today}
 
\begin{abstract}
We demonstrate that, for the case of quasi-equipartition between the
velocity and the magnetic field, the Lagrangian-averaged
magnetohydrodynamics $\alpha-$model (LAMHD) reproduces well both the
large-scale and small-scale properties of turbulent flows; in
particular, it displays no increased (super-filter) bottleneck effect
with its ensuing enhanced energy spectrum at the onset of the
sub-filter-scales.  This is in contrast to the case of the neutral
fluid in which the Lagrangian-averaged Navier-Stokes $\alpha-$model is
somewhat limited in its applications because of the formation of
spatial regions with no internal degrees of freedom and subsequent
contamination of super-filter-scale spectral properties.  \resp{}{We
  argue that, as the Lorentz force breaks the conservation of
  circulation and enables spectrally non-local energy transfer
  (associated to Alfv\'en waves), it is responsible for the absence of
  a viscous bottleneck in MHD, as compared to the fluid case.  As
  LAMHD preserves Alfv\'en waves and the circulation
  properties of MHD, there is also no (super-filter)
  bottleneck} found in LAMHD, making this method capable of large
reductions in required numerical degrees of freedom; specifically, we
find a reduction factor of $\approx 200$ when compared to a direct
numerical simulation on a large grid of $1536^3$ points at the same
Reynolds number.

\end{abstract}
 
\maketitle

\section{Introduction}

When large-scale numerical simulations of astrophysical or geophysical
magnetohydrodynamics (MHD) are desired, all dynamical scales of the
physical system are rarely, if ever, resolved.  For this reason,
sub-grid-scale (SGS) modeling of MHD dynamics in the context of computations in the geophysical and astrophysical context is required.
This modeling can be achieved implicitly, in the simplest example by employing a dissipative
numerical scheme, or it can be done explicitly by creating a Large Eddy Simulation
(LES--see \cite{MK00} for a recent review). Explicit methods for MHD are not as pervasive as they are in engineering, or for geophysical and atmospheric flows.  In fact,
modeling for MHD is a
relatively new field (see \cite{PFL76,Y87}).  One problem with
extending the LES methodology for hydrodynamic turbulence to MHD is that most LES are
based upon eddy-viscosity concepts \cite{MK00}, which can be related
to a known power law of the energy spectrum \cite{ChLe1981} (although generalizations can be devised, see e.g. \cite{BaPoPo+2008}), or upon
self-similarity.  For MHD, the underlying assumption of
 locality of interactions in Fourier
space is not necessarily valid \cite{AMP05a,MAP05} (a contradiction of
self-similarity) and spectral eddy-viscosity concepts \cite{ZSG02}
cannot be applied in a straightforward manner as neither kinetic nor magnetic energy is a
conserved quantity and the general expression of the energy spectrum
is not known at this time \cite{I64,K65,GoSr1995,GaNaNe+2000,MiPo2007a,PoMiMo+2008,MaCaBo2008}. 
Purely dissipative models
\cite{TFS94,AMK+01} are inadequate as they ignore the exchange of
energy at sub-filter scales between the velocity and magnetic fields
and such models have been shown to suppress small-scale dynamo action
\cite{HB06} and any inverse cascade from the sub-filter scales
\cite{MC02}.  \add{}{A satisfactory LES for MHD has been proposed} for the case starting
with some degree of alignment between the velocity and magnetic fields
\cite{LS91,MC02}.  Other restricted-case MHD-LES are applicable to low
magnetic Reynolds number \cite{PPP04,KM04,PMM+05}.  
Extensions of spectral models to MHD based on two-point closure formulations of the dynamical equations proposed recently look promising in the analysis of turbulent flows and of the dynamo mechanism \cite{BaPoPo+2008}.
Finally, though technically
not an LES, there are also hyper-resistive models for MHD which
require rescaling of the length (wavenumber) scales to a known direct
numerical simulation (DNS) \cite{HB06}.

One model which can be written as an LES is the Lagrangian-averaged
MHD (\lamhd) equations \cite{H02a, H02b, MP02}.  It has been shown to
reproduce a number of features of DNS. \resp{}{In two dimensions (2D)
  for Taylor Reynolds numbers ($R_\lambda$) up to $\approx5000$ it has
  been shown to reproduce selective decay, the inverse cascade of
  mean-square vector potential, and dynamic alignment between
  the velocity and magnetic fields \cite{MMP05a} as well as the statistics \vier{of} small-scale cancellation \cite{PGMP05} and intermittency
  \cite{PGHM+06}.  In three dimensions (3D) at Reynolds numbers ($Re$)
  of $\approx500$, \lamhda reproduced the inverse cascade of magnetic
  helicity (associated with the development of force-free magnetic
  field) and the helical dynamo effect \cite{MMP05b}.} \neu{It has
  also been tested (up to kinetic $Re\approx3000$, magnetic
  $Re\approx300$) for its ability to predict the critical magnetic
  Reynolds number for a non-helical dynamo at low magnetic Prandtl
  number \cite{Mi2006a}. \lamhda performed well in all these
  tests. Its equivalent hydrodynamic model, the Lagrangian-averaged
  Navier-Stokes (LANS) equations, also performed well in tests at
  $R_\lambda\lessapprox300$ (see \cite{CHO+05} and references in
  \cite{PGHM+07a}).  However, above $Re\approx3000$
  ($R_\lambda\approx800$),} it was shown that placing the filter width
in the inertial range leads to contamination of the super-filter-scale
properties (such as the spectra) \vier{for \lans.} We refer here to this effect as the
super-filter-scale bottleneck, which \resp{}{may be} different in
nature from the viscous bottleneck observed in some DNS of the
Navier-Stokes equations. The contamination may be linked to the
formation of spatial regions in the flow with no internal degrees of
freedom (so-called ``rigid bodies'') \cite{PGHM+07a}, which also
correspond to the development of a secondary inertial range of the
LANS equations at sub-filter scales. This \resp{}{super-filter-scale
  contamination} provides an effective constraint on the filter size
and, hence, on the available reduction of the total number of the
(numerical) degrees of freedom (\dof) needed to reproduce the
large-scale dynamics of the flow at a given Reynolds number; a factor
of $\approx 10$ \resp{}{can be achieved.  This limitation is not
  apparent} in low and moderate \resp{}{Reynolds number} (resolution)
simulations (e.g., $64^3$ \lansa compared with $256^3$ DNS) as the
scale separation is not enough for the above-mentioned phenomenon of
contamination of small-scale spectra because of rigid body regions in
the flow to appear. \resp{}{The bottleneck (and super-filter-scale
contamination) was not studied as such but neither was it observed
in 2D \lamhda for high Reynolds number \cite{MMP05a,PGMP05,PGHM+06}.
3D \lamhda has only been tested} at \add{}{more} moderate Reynolds
number \cite{MMP05b} (see also \cite{MiPoSu2008} for a recent review).
The aim of the present work is, thus, to \resp{}{determine if} \lamhda
in three space dimensions, \add{}{for higher Reynolds number}
\resp{}{develops problems similar to that of \lans.  Specifically, we
  test for the existence of} spatial regions with no available
internal degrees of freedom. We show in the following that \lamhda
behaves better in this respect than \lans, and, thus, continues to
appear as a promising model for MHD flows.

\section{The equations of motion}
\label{SEC:DETAILS}

We consider the incompressible MHD equations for a fluid with
constant density,
\begin{eqnarray} \partial_t\vec{v} + \boldsymbol{\omega} \times \vec{v}
 = \vec{j} \times \vec{b} - \boldsymbol{\nabla} p + \nu \nabla^2 \vec{v} \nonumber \\ 
\partial_t \vec{b} = \boldsymbol{\nabla} \times \left( \vec{v} \times \vec{b} \right) + \eta \nabla^2 \vec{b}
 \nonumber \\
 \boldsymbol{\nabla} \cdot \vec{v} =  \boldsymbol{\nabla} \cdot \vec{b} = 0,
\label{eq:mhd}
\end{eqnarray}
where $\vec{v}$ and $\vec{b}$ denote respectively the velocity and magnetic fields, $p$
the pressure divided by the density, $\nu$ the kinematic viscosity,
and $\eta$ the magnetic diffusivity.  
\resp{}{As is well known, in incompressible MHD, Alfv\'en waves will travel
  along a uniform background field, $\vec{b}_0$.  From linear
  perturbation analysis the dispersion relation between wavenumber,
  $k$, and frequency, $\omega$, is
\begin{equation}
\left(\omega+i\eta k^2\right)\left(\omega+i\nu k^2\right) = k^2b_0^2\,.
\end{equation}
The wave speed is $|\vec{b}_0|$ and, assuming $\eta=\nu$, the amplification
factor is given by $\exp(-\eta k^2t)$.}
The ideal ($\eta = \nu = 0$) quadratic invariants for MHD are in the
$L^2$ norm.  For example, the total energy is given by:
\begin{equation}
E_T = \frac{1}{2} \left( ||v||_2 + ||b||_2 \right)\equiv
\frac{1}{2} \frac{1}{D} \int_D \left(  |\vec{v}|^2 + |\vec{b}|^2 \right) d^3x.
\end{equation}

  The \lamhda equations \cite{H02a}
are given by
\begin{eqnarray} \partial_t\vec{v} + \boldsymbol{\omega} \times \vec{u}
 = \vec{j} \times \bar{\vec{b}} - \boldsymbol{\nabla} \Pi + \nu \nabla^2 \vec{v} \nonumber \\ 
\partial_t \bar{\vec{b}} = \boldsymbol{\nabla} \times \left( \vec{u} \times \bar{\vec{b}} \right) + \eta \nabla^2 \vec{b}
 \nonumber \\
 \boldsymbol{\nabla} \cdot \vec{v} =  \boldsymbol{\nabla} \cdot \vec{u} =  \boldsymbol{\nabla} \cdot \vec{b} =  \boldsymbol{\nabla} \cdot \bar{\vec{b}}  = 0,
\label{eq:lamhd}
\end{eqnarray}
where $\vec{u}$ ($\bar{\vec{b}}$) denotes the filtered component of
the velocity (magnetic) field and $\Pi$ the modified pressure.
Filtering is accomplished by the application of a normalized
convolution filter $L: f \mapsto \bar{f}$ where $f$ is any scalar or
vector field.  By convention, we define $\vec{u} \equiv
\bar{\vec{v}}$.  
\add{}{\lamhda in the form given in Eqs. (\ref{eq:lamhd}) is both computationally efficient and makes clear that Alfv\'en's theorem is preserved by the model: the smoothed magnetic field is advected by the smoothed velocity.}
In the remainder of this paper, we take $\eta=\nu$ (unit magnetic Prandtl number) and, thus, it is sufficient to introduce \add{}{the same filtering for the velocity and magnetic fields in this case.  This allows us to write \lamhda in
LES form,
\begin{eqnarray} \partial_t\vec{u} + \bar{\boldsymbol{\omega}} \times \vec{u}
 = \bar{\vec{j}} \times \bar{\vec{b}} - \boldsymbol{\nabla} \bar{\Pi} + \nu \nabla^2 \bar{\vec{v}} - \vec{\nabla}\cdot\bar{\tau} \nonumber \\ 
\partial_t \bar{\vec{b}} = \boldsymbol{\nabla} \times \left( \vec{u} \times \bar{\vec{b}} \right) + \eta \nabla^2 \bar{\vec{b}} - \vec{\nabla}\cdot\bar{\tau}^b \nonumber \\
 \boldsymbol{\nabla} \cdot \vec{v} =  \boldsymbol{\nabla} \cdot \vec{u} =  \boldsymbol{\nabla} \cdot \vec{b} =  \boldsymbol{\nabla} \cdot \bar{\vec{b}}  = 0.
\label{eq:lamhdLES}
\end{eqnarray}
} We choose as our filter the inverse of a Helmholtz
operator, $L = \mathcal H^{-1} = (1 - \alpha^2\nabla^2)^{-1}$.
Therefore, $\vec{u} = g_\alpha \otimes \vec{v}$ where $g_\alpha$ is
the Green's function for the Helmholtz operator, $g_\alpha(r) = \exp
(-r/\alpha)/(4\pi\alpha^2r)$ (i.e., the Yukawa potential), or in
Fourier space, $\hat{\vec{u}}(k) = \hat{\vec{v}}(k)/(1+\alpha^2k^2)$.
The effective filter width is, thus, approximately $\alpha$.
\add{}{With this choice,
the Reynolds (turbulent) SGS stress tensor is given by
\begin{equation}
\bar{\tau} = \alpha^2 \left( \vec{\nabla}\vec{u}\cdot\vec{\nabla}\vec{u}^T +
\vec{\nabla}\vec{u}\cdot\vec{\nabla}\vec{u} -
\vec{\nabla}\vec{u}^T\cdot\vec{\nabla}\vec{u} -
\vec{\nabla}\bar{\vec{b}}\cdot\vec{\nabla}\bar{\vec{b}}^T -
\vec{\nabla}\bar{\vec{b}}\cdot\vec{\nabla}\bar{\vec{b}} +
\vec{\nabla}\bar{\vec{b}}^T\cdot\vec{\nabla}\bar{\vec{b}}
 \right) 
\end{equation}
and the divergence of the electromotive-force (emf) SGS stress tensor by
\begin{equation}
 \vec{\nabla}\cdot\bar{\tau}^b = \eta\alpha^2 \nabla^4 \bar{\vec{b}}.
\label{eq:emf}
\end{equation}
In this form, the expression of the SGS tensors make explicit the fact that
$\vec{u} = \pm\bar{\vec{b}}$ Alfv\'en waves are preserved even in the subgrid scales.
\resp{}{These $\vec{u} = \pm\bar{\vec{b}}$ waves} \resp{}{travel along $\bar{\vec{b}}_0$ (the smoothed
and unsmoothed fields are identical for uniform ${\vec{b}}_0$) and the dispersion relation is
\begin{equation}
\left(\omega+i\nu k^2\right)\left(\omega+i\eta k^2 (1+\alpha_M^2k^2)\right) = k^2\bar{b}_0^2 \frac{1+\alpha_M^2k^2}{1+\alpha_K^2k^2}\,,
\end{equation}
where $\alpha_K$ and $\alpha_M$ are the filter widths for the
smoothing of the velocity and magnetic fields, respectively.  For
$\alpha\equiv\alpha_K=\alpha_M$ \vier{and $\eta=\nu$} (the case we study), the wave speed is
given by $\bar{\vec{b}}_0 \vier{\left(1-(\eta k\alpha^2k^2/\bar{b}_0)^2/8 + \mathcal{O}((\eta k\alpha^2k^2/\bar{b}_0)^6)\right)}$, the strength of the smoothed background
magnetic field \vier{reduced by an order $\alpha^4k^4$ term.}  The amplification factor is given by
 \vier{$\exp\left(-\eta k^2t (1+\alpha^2k^2/2)\right)$ for both } $u=-\bar{b}$ waves
traveling in the direction of $\bar{\vec{b}}_0$ and
$u=\bar{b}$ waves traveling anti-parallel to
$\bar{\vec{b}}_0$.}  Finally,} the ideal quadratic invariants for
\lamhda are in the $H^1_\alpha(\bar{f})$ norm.  For example, the total
energy is given by a mixture of the smooth and rough fields, namely:

\begin{eqnarray}
E_T^\alpha = \frac{1}{2} \left( ||u||_2^\alpha + ||\bar{b}||_2^\alpha \right) \equiv
\frac{1}{2} \frac{1}{D} \int_D \left( \vec{u} - \alpha^2\nabla^2 \vec{u}\right)
\cdot \vec{u} + \left( \bar{\vec{b}} - \alpha^2\nabla^2 \bar{\vec{b}}\right)
\cdot \bar{\vec{b}}~d^3x \nonumber \\
= \frac{1}{2} \frac{1}{D} \int_D \vec{v}\cdot\vec{u} + \vec{b}\cdot\bar{\vec{b}}~d^3x.
\end{eqnarray}

We solve both sets of equations, Eqs. (\ref{eq:mhd}) and
(\ref{eq:lamhd}), for one specific instance of a decaying MHD flow,
using a parallel pseudospectral code \cite{GMD05,GMD05b} in a
three-dimensional (3D) cube with periodic boundary conditions.  The
initial conditions for the velocity and magnetic fields are
constructed from a superposition of three Beltrami (helical) ABC flows
to which smaller-scale random fluctuations are added with initial
kinetic and magnetic energy $E_K = E_M = 0.5$, magnetic helicity $H_M
= <\vec{a}\cdot\vec{b}> \approx 0.45$ ($\vec{b}
=\vec{\nabla}\times\vec{a}$ where $\vec{a}$ is the vector potential
and the brackets denote volume average), and the initial co-alignment
of the fields, $\left<\vec{v}\cdot \vec{b}\right>
\left<|\vec{v}||\vec{b}|\right>^{-1} \approx 10^{-4}$ (see
\cite{MiPoMo2006,MiPo2007a} for details).  A MHD-DNS with a resolution
$N^3 = 1536^3$ \add{}{(i.e., 1536 grid points in real space in each
  direction)} and $\eta = \nu = 2 \cdot 10^{-4}$ is used as our high
Reynolds number test case for the \lamhda model.  The DNS computation
is stopped when the growth of the total dissipation begins to enter
the saturation phase ($t=3.7$), at which time the Reynolds number
based on the mechanical integral scale is $Re \approx 9200$ and the
Taylor Reynolds number $\approx 1100$.  The MHD flow resulting from
the initial conditions employed has previously been analyzed for its
spectral properties and for the spatial structures it develops
\cite{MiPoMo2006,MiPo2007a,MaPoMi+2008}.  In this paper, we perform a
simulation with similar initial conditions and parameters but now
using \lamhda at a resolution of $512^3$ grid points; we also perform
for comparison purposes a Navier-Stokes \lansa run with the same
initial velocity field but with $\vec{b} \equiv 0$, on a grid of
$512^3$ points. In both cases, the filter width is $\alpha=2\pi/18$
($k_{\alpha}=18$) and is, thus, large enough to preclude any artifact
of numerical resolution altering the results.  Based on previous
analyses \cite{FHT01,PGHM+07a}, we estimate ${k_{max}}/{k_\eta^\alpha}
\approx 2.4$ (where $k_{max}$ is the maximum wavenumber resolved in
the simulation, and $k_\eta^\alpha$ is the \lamhda dissipation scale)
using computations conducted for $\eta=\nu=6\cdot10^{-4}$ with a
Reynolds number of $Re \approx 2200$.  \resp{}{However, the main point
  of using such a large filter is to test if \lamhda fails in the same
  way as \lans.}  We finally perform a LES simulation in a $256^3$
grid using the \lamhda equations with the same viscosity and
diffusivity as the $1536^3$ DNS used for the comparison.  \resp{}{In
  this way, we extend the $Re\approx9200$ computation in time by a
  factor of 3.}

\section{Results}
\subsection{Spectral contamination in \lansa \ for an ABC flow and its absence in the MHD case} 
\label{SEC:LAMHDGOOD}

\begin{figure}[htbp]\begin{center}\leavevmode \centerline{%
  \begin{tabular}{c@{\hspace{.15in}}c}
  \includegraphics[width=8.95cm]{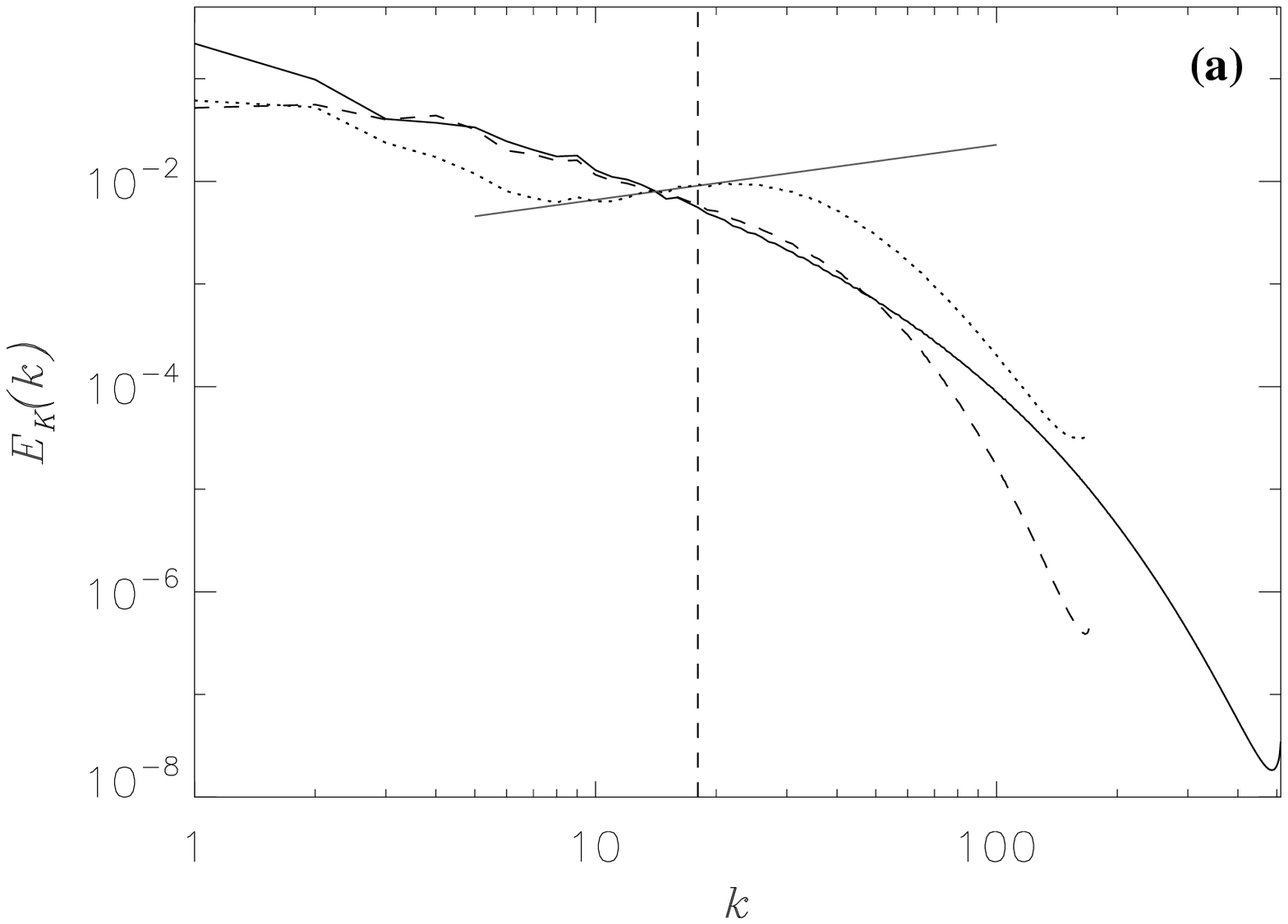} &
  \includegraphics[width=8.95cm]{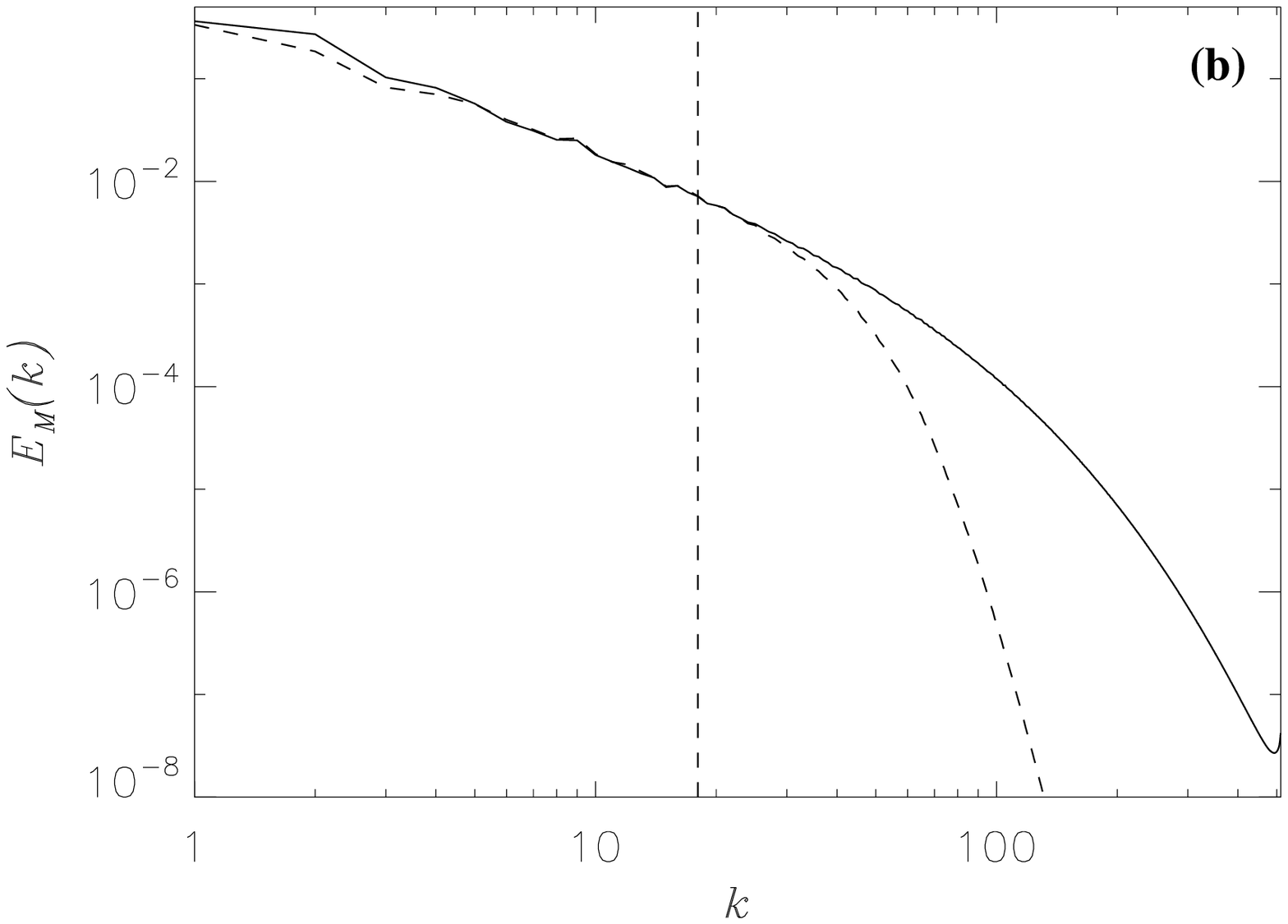}  \end{tabular}}
  \caption{
  {\bf(a)} Spectra of kinetic energy \AD{(normalized to DNS $E_K(14)$, see text)} for $1536^3$ MHD DNS (solid line), $512^3$ \lamhda (dashed), and $512^3$ \lansa (dotted), in the latter case with ${\bf b}\equiv 0$ at all times but otherwise identical conditions. For intermediate scales, $k \in [5,40]$, \lamhda reproduces the
    scaling of the DNS, the larger scales being affected by slight differences
    in initial conditions, see text.  For $k$ close to the filter scale ($k \in [k_\alpha/2,k_\alpha]$), a
    positive power law, $k^{0.5}$ (gray line), is found for \lans.
{\bf(b)} Spectra of magnetic energy \AD{(normalized to DNS $E_M(14)$)} for the same runs: \lamhda reproduces the
    scaling of the DNS even beyond the filter wavenumber, $k_\alpha = 18$ as indicated by the vertical dashed line. \lamhda exhibits neither the positive
    power-law nor the super-filter-scale spectral contamination
    associated with high Reynolds number \AD{\lansa} modeling seen in (a).  }
  \label{FIG:COMP1} \end{center} \end{figure}

One of the main findings of our preceding work with \lansa on the Navier-Stokes equations is that a $k^{+1}$ scaling develops in the (kinetic) energy spectrum at sub-filter scales; this leads to a contamination of super-filter scales because of detailed energy conservation (per triadic interactions).
This \lansa $k^{+1}$ spectrum (together with super-filter-scale spectral
contamination) has only recently been recognized, in the case of one specific forcing function
at large Reynolds number \cite{PGHM+07a}, but such a spectral contamination
has not yet been generally demonstrated (although theoretical arguments for the $k^{+1}$  spectrum have been given in \cite{PGHM+07a}).  Thus, we first confirm
its presence in a \lansa simulation with the same viscosity and the
(nearly) same initial conditions for the velocity field as for the
MHD-DNS (and \lamhda runs) examined in this paper, and based on large-scale ABC flows with superimposed random noise at small scale.  
Due to the presence of random noise and considering the differences in resolution and the presence of a filter in the \lamhda runs, the initial conditions were not exactly reproduced, although the same procedure was used to generate them.
In the present Navier-Stokes case, we find again what can be called an enhanced (super-filter-scale) bottleneck:
 the positive power-law spectral
contamination of the kinetic energy spectrum $E_K(k)$ in the \lansa run is observed for times after the peak of dissipation (see dotted line,
Fig. \ref{FIG:COMP1}a).  The fitted spectrum is $k^{+0.5}$ (note that
$k^{+1}$ requires the entire LANS spectrum to be resolved, and therefore has only been observed for much larger values of
${k_{max}}/{k_\eta^\alpha}$).

\resp{}{For the given parameters and initial conditions, we find the
super-filter-scale bottleneck for \lans.} 
However, when integrating the MHD equations with the Lagrangian model (dashed line, Fig. \ref{FIG:COMP1}a)  \resp{}{with these same parameters,} no such contamination is present.
Note that the spectra for the DNS-MHD
are shown at the time of peak dissipation, while the spectra for the
Lagrangian-averaged models are for a slightly later time in order to
allow for the possible formation of rigid bodies, which are known to be at the source of the spectral contamination close to the filter wavenumber in the Navier-Stokes case.
\add{}{For this reason, and due to the slight differences in initial conditions,
we have chosen to plot spectra normalized to that of the DNS at $k=14$ to emphasize the scaling.} 
For most of
the inertial range \add{}{(also in an approximate sense below the filter
width $\alpha$) the scaling of $E_K(k)$ is reproduced by the \lamhda
simulation.  The sub-filter scaling for \lamhda is not as steep as MHD, but is not a positive scaling law.}  The agreement for $E_M(k)$ is remarkable.  More
importantly, neither positive-power-law spectra nor contamination of
the super-filter-scale spectra are evidenced at all.

\subsection{The lack  of rigid bodies in \lamhda \ in the large$-\alpha$ limit for unforced flows}

\resp{}{Evidence for the development of rigid bodies in \lansa (which led
to its limited use as a LES) has only been shown for $l\ll\alpha$ \cite{PGHM+07a}.}
\add{}{Since investigation of the large$-\alpha$ limit
is not as computationally demanding \resp{}{ as the small$-l$ limit,} it is interesting to look at this
limit as a rough indication of what occurs for small $\alpha$ and
\resp{}{smaller $l$.}  This approach has been employed both for}
 the \lansa Navier-Stokes case in two dimensions \cite{LKT+07} and \add{}{in three dimensions} \cite{PGHM+07a}. 
In such a case, the purpose is to examine the properties of the model itself, as opposed to trying to reproduce large-scale properties, the large-scale behavior being reduced to a very small span of wavenumbers.
With this practice, the properties of the sub-filter-scales can be studied, to better understand the origin (or lack) of super-filter-scale contamination.  We now use this
limit to further explore the differences between \lamhda and \lans.
We employ simulations for the two models with the same initial conditions as
before, with $\eta = \nu = 5 \cdot 10^{-5}$ ($Re \approx 26,000$ at peak
of dissipation for \lamhd), and a resolution of $256^3$ grid points. 
Note that these dissipative coefficients are four times smaller than what was considered in the previous section since, for a fixed resolution, the achievable Reynolds number goes as $\alpha^{2/3}$.
\add{}{This follows for \lansa from the predicted (and verified) degrees of
freedom, $N^\alpha_{dof} \propto \alpha^{-1} Re^{3/2}$ \cite{FHT01,PGHM+07a}.  The scaling of \lamhda may differ, but the same value of the viscosity is employed
for the two models, regardless.}

\begin{figure}[htbp]\begin{center}\leavevmode \centerline{%
\includegraphics[width=8.95cm]{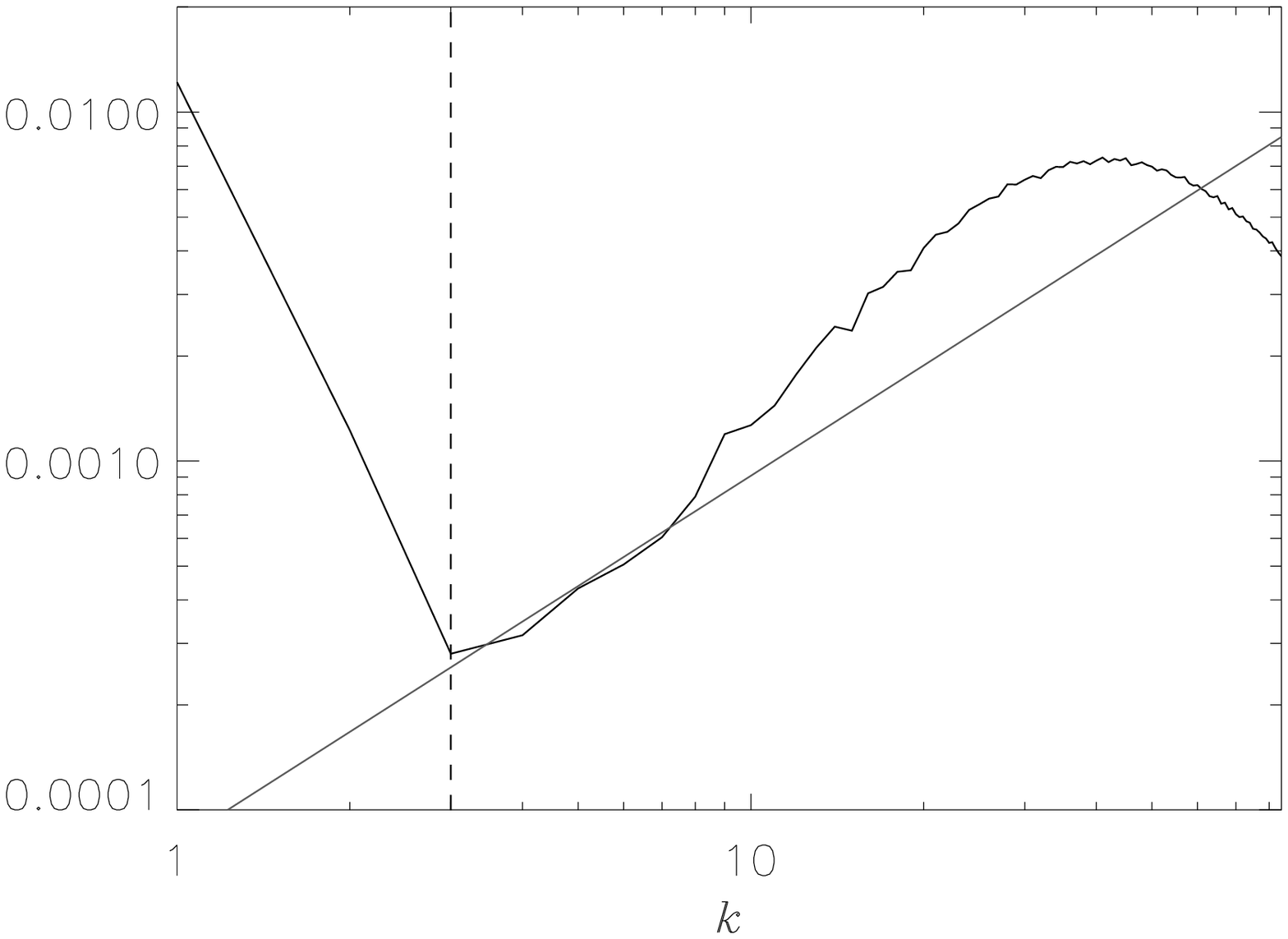}  }
  \caption{Spectrum of kinetic energy for a $256^3$ grid with $k_\alpha=3$ ($\nu =
    5\cdot10^{-5}$) \lans, ${\bf b}\equiv 0$ (Navier-Stokes case).  The fitted grey
    line, $k^{+1.1\pm.4}$, agrees with the rigid-body hypothesis for
    the inertial range \cite{PGHM+07a}.  This slope is followed by a
    steeper slope attributed to a bottleneck, with $k^{+1.5\pm.2}$.}
  \label{FIG:MASK} \end{center} \end{figure}

\begin{figure}[htbp]\begin{center}\leavevmode \centerline{%
\includegraphics[width=8.95cm]{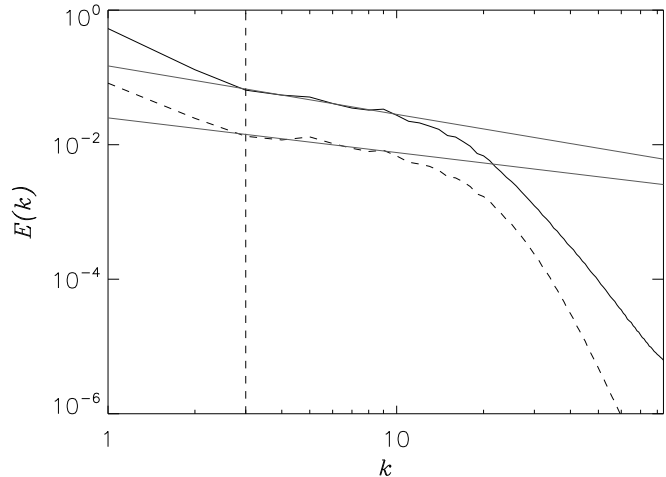} 
}
  \caption{Spectra for a $256^3$ grid with $k_\alpha=3$ ($\eta=\nu=5\cdot10^{-5}$)
    \lamhd, $Re \approx 26,000$: Total energy, $E_T(k)$, (solid line)
    and cross helicity, $H_C(k)$, (dashed).  The fitted slopes,
    $E_T(k)\sim k^{-0.7\pm.3}$ and $H_C (k)\sim k^{-0.5\pm.4}$ could agree
    with either Kolmogorov or IK predictions for \lamhda (see text) at this
    level of uncertainty.}
      \label{FIG:LARGEASPEC} \end{center} \end{figure}

For \lans, we observe
the expected $k^{+1}$ zero flux inertial range (see
Fig. \ref{FIG:MASK}) which is followed by a viscous (sub-filter-scale) bottleneck feature,
$k^{+1.5\pm.2}$, before the dissipative range proper.  We conducted a second
simulation with $\nu = 10^{-4}$ and found a $k^{1.4\pm.3}$ spectrum.
  This is analogous to results for DNS of the Navier-Stokes equations where
only the viscous bottleneck is observed at moderate Reynolds number and is preceded by
an inertial range only for higher Reynolds. These viscous bottlenecks \resp{}{may be} different in nature from the (super-filter-scale) bottlenecks discussed before, which are not associated to the onset of the dissipative range but to the development of a secondary inertial range in LANS below the filtering length, and result in contamination of the large (resolved) scales when the LANS equations are used as an LES.
Having confirmed that our analysis from the forced \lansa case extends
to the decaying \lansa simulation, we now apply it to \lamhd.  The
large$-\alpha$ \lamhda spectra are given in
Fig. \ref{FIG:LARGEASPEC}.  Notably, there is no positive-power-law
spectrum.

Predictions of energy spectra in the inertial range follow from  the global scaling laws for third-order structure functions
for isotropic, homogeneous turbulence.  Exact results for these structure functions have been \resp{}{found} for incompressible MHD \cite{PP98b}
and for \lamhda \cite{PGHM+06}.  The latter are, in terms of both the smooth fields \vier{${\bar {\bf z}}^{\pm} \equiv \vec{u}\pm\bar{\vec{b}}$} and the rough fields $\vec{z}^\pm \equiv \vec{v}\pm\vec{b}$ (where the z-fields are called the Els\"asser
variables):
\begin{equation} \left< \delta\bar{z}_\|^\mp(\vec{l})  \delta\bar{z}_i^\pm(\vec{l}) 
  \delta{z}_i^\pm(\vec{l}) \right> \sim \varepsilon_\pm^\alpha l \ ,  \label{eq:khlamhd} \end{equation}
where $\left< . \right>$ denotes volume averaging, $\delta
f(\vec{l}) \equiv f(\vec{x}+\vec{l}) - f(\vec{x})$, and $\delta
f_\|(\vec{l}) \equiv \left[\vec{f}(\vec{x}+\vec{l}) -
  \vec{f}(\vec{x})\right] \cdot \vec{l}$.  For sub-filter scales ($l
\ll \alpha$), $\bar{z}^\pm \sim l^2\alpha^{-2}{z}^\pm$ and the scaling law becomes dimensionally
$\bar{z}z\bar{z} \sim \varepsilon l$.  This implies a sub-filter scale
spectrum
corresponding to the invariants $E_{\pm}^\alpha \equiv ||\bar{\vec{z}}^\pm||^2_\alpha/2$ for the ideal non-dissipative case. We then have
 $E^\alpha_\pm(l)k \sim z^\pm\bar{z}^\pm \sim
(\varepsilon_\pm^\alpha)^{2/3}\alpha^{2/3}$ or, equivalently,
\begin{equation}
E^\alpha_\pm(k) \sim (\varepsilon_\pm^\alpha)^{2/3}\alpha^{2/3} k^{-1}
\label{eq:speclamhd}
\end{equation}
as for \lansa \cite{FHT01}.
Recall that in the flux relation,
Eq. (\ref{eq:khlamhd}), $\varepsilon_\pm^\alpha$ stands for the energy
transfer and dissipation rate of $E_\pm^\alpha$. 
Hence, the prediction, 
Eq. (\ref{eq:speclamhd}), for the spectra, $E^\alpha_\pm(k)$, is,
equivalently for $E_T^\alpha \equiv (||u||^2_\alpha+||b||^2_\alpha)/2$ and for $H_C^\alpha
\equiv \frac{1}{2} \frac{1}{D} \int_D \vec{v} \cdot \bar{\vec{b}}~d^3x$.  The spectra shown in
Fig. \ref{FIG:LARGEASPEC} for large$-\alpha$ \lamhda do not exclude,
due to the large uncertainties of the fitted power laws, the predicted
$k^{-1}$ spectra.  

A spectral prediction for \lamhda can also be arrived at by
dimensional analysis of the spectrum which follows the scaling ideas
originally due to Kraichnan \cite{K67} and which is developed
for \lansa in Ref. \cite{CHT05}.  Here, the energy dissipation
rate, \add{}{$\varepsilon^\alpha_\pm = dE^\alpha_\pm/dt$,} is related to the spectral energy
density by
\begin{equation}
\varepsilon^\alpha_\pm \sim (\mathfrak{t}_k)^{-1} \int E^\alpha_\pm(k)
\label{eq:scaling} 
\end{equation}
where $\mathfrak{t}_k$ is the turnover time for an eddy of size
$\sim k^{-1}$.  This turnover time is related to a ``velocity,''
$\bar{Z}_k^\pm$, (i.e., $\mathfrak{t}_k \sim 1 / (k\bar{Z}_k^\pm)$), where
$(\bar{Z}_k^\pm)^2 \sim \bar{Z}_k^\pm{Z}_k^\pm/(1+\alpha^2k^2) \sim k
E^\alpha_\pm(k)/(1+\alpha^2k^2)$.  Substitution into
Eq. (\ref{eq:scaling}) yields,
\begin{equation}
E^\alpha_\pm(k)
\sim (\varepsilon^\alpha_\pm)^{2/3} k^{-5/3}(1+\alpha^2k^2)^{1/3}
\label{eq:oldK41}
\end{equation}
or, for $\alpha k \gg 1$,
\add{}{\begin{equation}
E^\alpha_\pm(k)
\sim (\varepsilon^\alpha_\pm\alpha)^{2/3} k^{-1}.
\end{equation}}
In the Iroshnikov-Kraichnan \cite{I64,K65} (hereafter, IK) phenomenology, Alfv\'en waves 
(corresponding to either $z^\mp = 0$)
can only
interact nonlinearly when they collide along field lines (along which they travel
in opposite directions).  The characteristic time for an Alfv\'en wave
is $\mathfrak{t}_A \sim (kB_0)^{-1}$.  If this is less than
$\mathfrak{t}_k$, the effective transfer time $\mathfrak{t}_T$ is
increased, $\mathfrak{t}_T \sim \mathfrak{t}_k^2 /
\mathfrak{t}_A$. Substitution of this new transfer time into
Eq. (\ref{eq:scaling}) yields, \add{}{instead of Eq. (\ref{eq:oldK41})
\begin{equation}
E^\alpha_\pm(k)
\sim (\varepsilon^\alpha_\pm B_0)^{1/2} k^{-3/2}(1+\alpha^2k^2)^{1/2}
\end{equation}
or, for $\alpha k \gg 1$,
\begin{equation}
E^\alpha_\pm(k)
\sim (\varepsilon^\alpha_\pm B_0)^{1/2} \alpha k^{-1/2}.
\label{eq:IK}
\end{equation}
} The spectra shown in Fig. \ref{FIG:LARGEASPEC} for large$-\alpha$
\lamhda also agree with the IK predicted spectra, Eq. (\ref{eq:IK}).
In fact, the spectra more closely correspond to this prediction;
this is consistent with the fact that, for this flow, an IK spectrum $E(k)\sim k^{-3/2}$ is observed at large scale (followed by a weak turbulence anisotropic spectrum $E(k_{\perp})\sim k_{\perp}^{-2}$ at small scale) \cite{MiPo2007a}.
 Again, simulations at higher resolution are needed for a definite answer and
the result may not be universal 
as shown for example in the context of reduced MHD dynamics due to the presence of a strong uniform magnetic field 
${\bf B}_0$ \cite{DmGoMa2003} or for MHD with a strong ${\bf B}_0$ \cite{MaCaBo2008}.

\begin{figure}[htbp]\begin{center}\leavevmode
\centerline{%
\includegraphics[width=8.95cm]{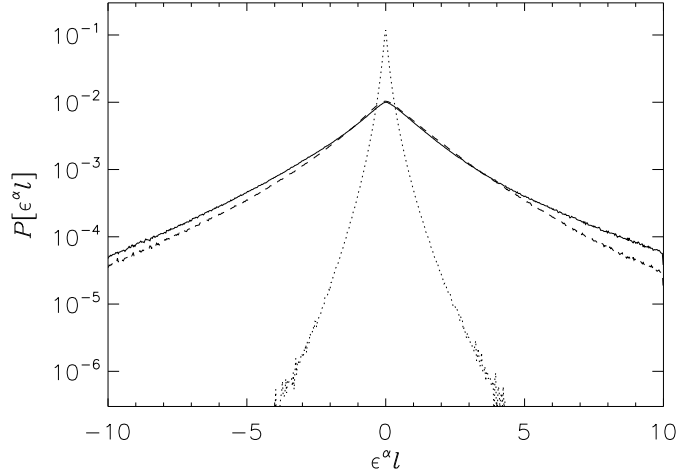} 
}
  \caption{PDFs of cubed increments. \AD{The cubed increments when
      averaged are equal to flux times length, $\varepsilon^\alpha
      \cdot l$.}  Here $l = 0.88\alpha$ ($\alpha=2\pi/3$).  The dotted
    line is \AD{$\delta u_\|(l) \delta u_i(l) \delta v_i(l)$} for
    \lans, solid for \lamhda \AD{$\delta\bar{z}_\|^-(\vec{l})
      \delta\bar{z}_i^+(\vec{l}) \delta{z}_i^+(\vec{l})$}, and dashed
    for \lamhda \AD{$\delta\bar{z}_\|^+(\vec{l})
      \delta\bar{z}_i^-(\vec{l}) \delta{z}_i^-(\vec{l})$.}  More of
    the volume gives no contribution to the flux for \lansa than for
    \lamhd, indicating no rigid bodies in \lamhd.  }
  \label{FIG:PDFS}
\end{center}
\end{figure}

Another indication of the zero-flux regions in \lansa is \resp{}{found} by
examining the spatial variation of the cubed increments associated
with the scaling laws
$\delta u_\|(l) \delta u_i(l) \delta v_i(l)$ for \lansa and
$\delta\bar{z}_\|^\mp(\vec{l}) \delta\bar{z}_i^\pm(\vec{l})
\delta{z}_i^\pm(\vec{l})$ for \lamhda
(note that one can transform this relation into the $u,\ v,\ b,\ {\bar b}$ variables).
\add{}{For a given length $l$, these cubed increments when averaged are related
with the energy fluxes by Eq. (\ref{eq:khlamhd}) (the \lansa relation
and the hydrodynamic and MHD relations are contained in this expression in the corresponding limits).  As a result of this correspondence, for brevity we will
indicate cubed increments in the figures as the corresponding energy flux times
the length used to compute the increments.  This also allows us to identify regions with zero cubed increments as rigid bodies (a rigid rotation has zero longitudinal increments).}
  Probability distribution
functions (PDFs), see Fig. \ref{FIG:PDFS}, indicate that \lamhda has a
much smaller proportion of its volume, which could potentially be rigid
bodies (i.e., frozen regions with no internal degrees of freedom \add{}{(zero velocity increment)}, which therefore do not contribute to the energy flux).  That is, more of the volume is contributing to the turbulent
cascade.  Snapshots for constructing the PDFs are taken from both $\alpha=2\pi/3$
Lagrangian-averaged models for times shortly after the peak of
dissipation and when the \lansa total dissipation is nearly equal to
that of \lamhd.  The strengths of the central peaks of the PDFs for
large$-\alpha$ are another indication that \lamhda inherits none of
the rigid-body or zero-flux-region problems of \lans.

\subsection{Why are spectral properties of \lamhd\  better than in the fluid case?}

\begin{figure}[htbp]\begin{center}\leavevmode
\centerline{%
  \includegraphics[width=8.95cm]{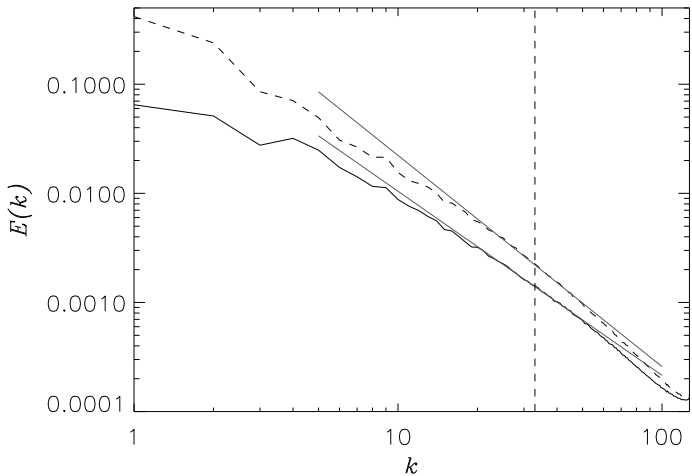}
}
  \caption{Spectra for a $384^3$ grid with $k_\alpha=33$ obtained from the modified-\lamhd \ (see text) shortly after the maximum of dissipation: kinetic
    energy (solid) and magnetic energy (dashed); the \lamhda
    equations have been modified by removing the hyperdiffusive
    turbulent emf.  Even without hyperdiffusivity, no positive
    power-law is found.  Instead, fits (grey lines) for kinetic and
    magnetic energy spectra near the filtering length are $k^{-1.7\pm.1}$ 
    and $k^{-1.9\pm.1}$, respectively.}
  \label{FIG:HYPER}
\end{center}
\end{figure}

Why does \lamhda not exhibit the same spectral contamination as \lans?
\resp{}{One possible cause is the
hyperdiffusivity term
seen in the LES form for \lamhd,
Eq. (\ref{eq:emf}),
whereas} there is no hyperviscosity-like term in \lans.  To test if this hyperdiffusion
is responsible for the lack of spectral contamination in \lamhd, we
removed the
hyperdiffusion by \add{}{setting $\bar{\tau}^b=0$ in Eqs. (\ref{eq:lamhdLES}) or,
equivalently, by} substituting $\eta \nabla^2\bar{\vec{b}}$ for $\eta
\nabla^2\vec{b}$ in Eqs. (\ref{eq:lamhd}).
We then start the run from the same initial conditions but now with these new equations \resp{}{employing} $\alpha = 2\pi/33$ \add{}{and} $\nu=\eta=2\cdot10^{-4}$ at a
resolution of $384^3$ (with hyperdiffusion, a smaller resolution of
$256^3$ is possible, see Section \ref{SEC:LES}).
 Note that
such a modified \lamhda model is not expected to, nor found to, perform well as a SGS model; 
this numerical experiment is performed here only in order to assess the effect of the hyper-diffusive term introduced by the $\alpha$ modeling.
 We find that hyperdiffusion
is {\sl not} responsible for the lack of a $k^{+1}$ spectral
contamination in \lamhda (see Fig. \ref{FIG:HYPER}).

\resp{}{Other possible causes for \lamhda not exhibiting the
  super-filter-scale bottleneck as does \lansa are the} actual
physical differences between the two fluids \resp{}{that are modeled,}
Navier-Stokes and MHD.  First, unlike incompressible Navier-Stokes,
MHD supports oscillatory solutions (Alfv\'en waves) which are linked
to enhanced spectral nonlocality of energy transfer \cite{AMP05a,
  Al2007a} leading to dynamic interactions between widely separated
scales.  \resp{}{ For Navier-Stokes, the depletion of energy transfer
  due to local interactions at some cutoff in wavenumber is believed
  to bring about the bottleneck effect
  \cite{HSL+82,LMG95,MCD+97,MAP06}.  However, related to the
  spectrally nonlocal energy transfer via Alfv\'en waves,} MHD does
not seem to exhibit a bottleneck in its spectra between the inertial
and dissipative ranges \cite{MiPo2007a}.  \resp{}{As \lamhda supports
  Alfv\'en waves at all scales (and alters their dissipation \vier{and wave speed}
  appreciably only for sub-filter scales), the same physics could be
  behind the lack of a super-filter-scale bottleneck in \lamhd.}
  
\begin{figure}[htbp]\begin{center}\leavevmode \centerline{%
 \includegraphics[width=8.95cm]{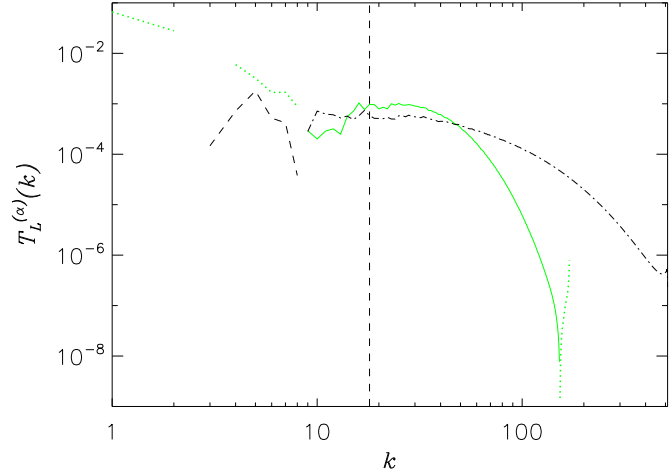}}
  \caption{(Color online) Spectral transfer due to the Lorentz force, \neu{$T_L$ (for $1546^3$ DNS)} and  $T_L^\alpha$ (for $512^3$ $k_\alpha=18$
    \lamhd) \neu{at a time just prior to the peak of dissipation.}  \neu{Positive $T_L$ is shown as dash-dotted lines and
    negative $T_L$ as dashed lines.}  Positive $T_L^\alpha$ is shown as solid (green online) lines and
    negative $T_L^\alpha$ as dotted (green online) lines. \neu{\lamhda qualitatively reproduces the transfer of kinetic energy in MHD.}}
  \label{FIG:TRANSFER} \end{center} \end{figure}

Another difference between the fluid and MHD cases is the geometry of
the dissipative structures: one finds vortex filaments for
Navier-Stokes at high value of the vorticity, and current and
vorticity sheets for MHD; sheets which are found to roll-up at high
Reynolds number \cite{MiPoMo2006}.  \resp{}{It has been claimed that
  the development of helical filaments in the fluid case can lead to
  the depletion of nonlinearity and the quenching of local
  interactions \cite{MT92,T01} and, hence, to the viscous bottleneck.
  A similar energy transfer depletion may occur in \lans.}  In
\cite{PGHM+07a} \resp{}{evidence is presented} that Taylor's frozen-in turbulence
hypothesis applied to Lagrangian averages leads to the formation of
``rigid bodies'' in the flow wherein there are no internal degrees of
freedom and no transfer of energy to smaller scales (i.e. regions with
$\varepsilon \sim \delta u_\|^3/l = 0$ as well as
$\boldsymbol{\omega}\times\vec{v} = 0$).  These regions are likely
related to the shorter, thicker vortex filaments formed and the
suppression of vortex stretching dynamics as $\alpha$ is increased
\cite{CHM+99}.  As MHD has spectrally non-local transfer (e.g.,
velocity at large scales does stretching of magnetic field lines at
small scales) \resp{}{this leads to the break up of these rigid bodies
in the \lamhda case and the breakup of the viscous bottleneck
  in the MHD case.  The magnetic field} interaction with the large
scale \resp{}{velocity can} re-enable transfer of energy to smaller
scales \resp{}{of the velocity field.}  Indeed, defining the kinetic
spectral transfer due to the Lorentz force as
\begin{equation}
T^\alpha_L(k) \equiv \int \hat{\vec{u}}_k \cdot \left( \widehat{\vec{j} \times \bar{\vec{b}}} \right)^*_k d\Omega_k
\label{eq:transfer_alpha}
\end{equation}
\resp{}{for \lamhda, and as
\begin{equation}
T_L(k) \equiv \int \hat{\vec{v}}_k \cdot \left( \widehat{\vec{j} \times {\vec{b}}} \right)^*_k d\Omega_k
\label{eq:transfer}
\end{equation}
for MHD,} we see in Fig. \ref{FIG:TRANSFER} that the Lorentz force is
removing large-scale kinetic energy and supplying small-scale kinetic
energy; this \resp{}{effectively bypasses} the formation of rigid bodies
\resp{}{for \lamhda and the viscous bottleneck for MHD (note that
  Eqs. (\ref{eq:transfer_alpha}) and (\ref{eq:transfer}) do not detail
  the scales at which magnetic energy is created or destroyed).}

This \add{}{argument can also be recast} in terms of Kelvin's circulation theorem. \resp{}{For Navier-Stokes,} the circulation $\Gamma$ of the velocity \resp{}{$\vec{v}$} is conserved in the ideal case for barotropic flows. In ideal \resp{}{MHD,} this conservation is broken by the Lorentz force,
\resp{}{\begin{equation}
\frac{d \Gamma}{dt} = \frac{d}{dt} \oint_{\cal C} \vec{v}\cdot d\vec{r} 
    = \oint_{\cal C} \vec{j} \times {\vec{b}} \cdot d\vec{r},
\label{eq:kelvin}
\end{equation}
} where ${\cal C}$ is any material curve. As a result, while in ideal \resp{}{Navier-Stokes} a material curve ${\cal C}$ defines the boundary of a vorticity tube with fixed strength, in \resp{}{MHD} these structures are deformed and \resp{}{their} vorticity content changed by the Lorentz force.
\resp{}{A similar result follows for \lamhda and \lans,}
\begin{equation}
\frac{d \Gamma}{dt} = \frac{d}{dt} \oint_{\cal C} \vec{u}\cdot d\vec{r} 
    = \oint_{\cal C} \vec{j} \times \bar{\vec{b}} \cdot d\vec{r}\,.
\end{equation}
\resp{}{Breaking the conservation of circulation in this way can
  prevent the formation of a bottleneck.  For example, for the fluid
  case in the Clark$-\alpha$ model (which differs from \lansa only in
  the conservation of $\Gamma$), it was also found that no
  super-filter-scale bottleneck was present \cite{PiGrHoMi+2008}.}

\subsection{LES Application}
\label{SEC:LES}

\begin{figure}[htbp]\begin{center}\leavevmode
\centerline{%
  \begin{tabular}{c@{\hspace{.15in}}c}
  \includegraphics[width=8.95cm]{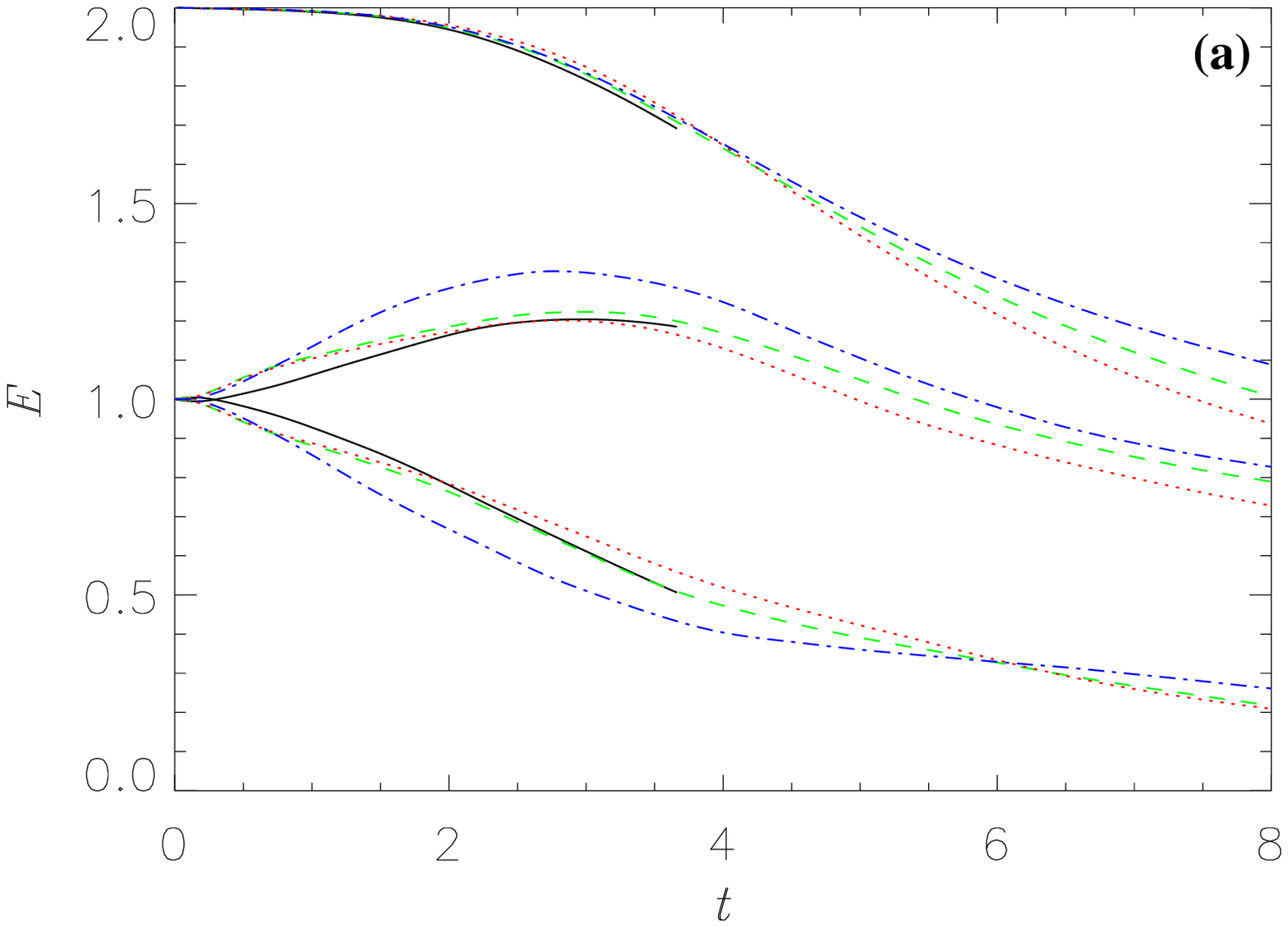} &
  \includegraphics[width=8.95cm]{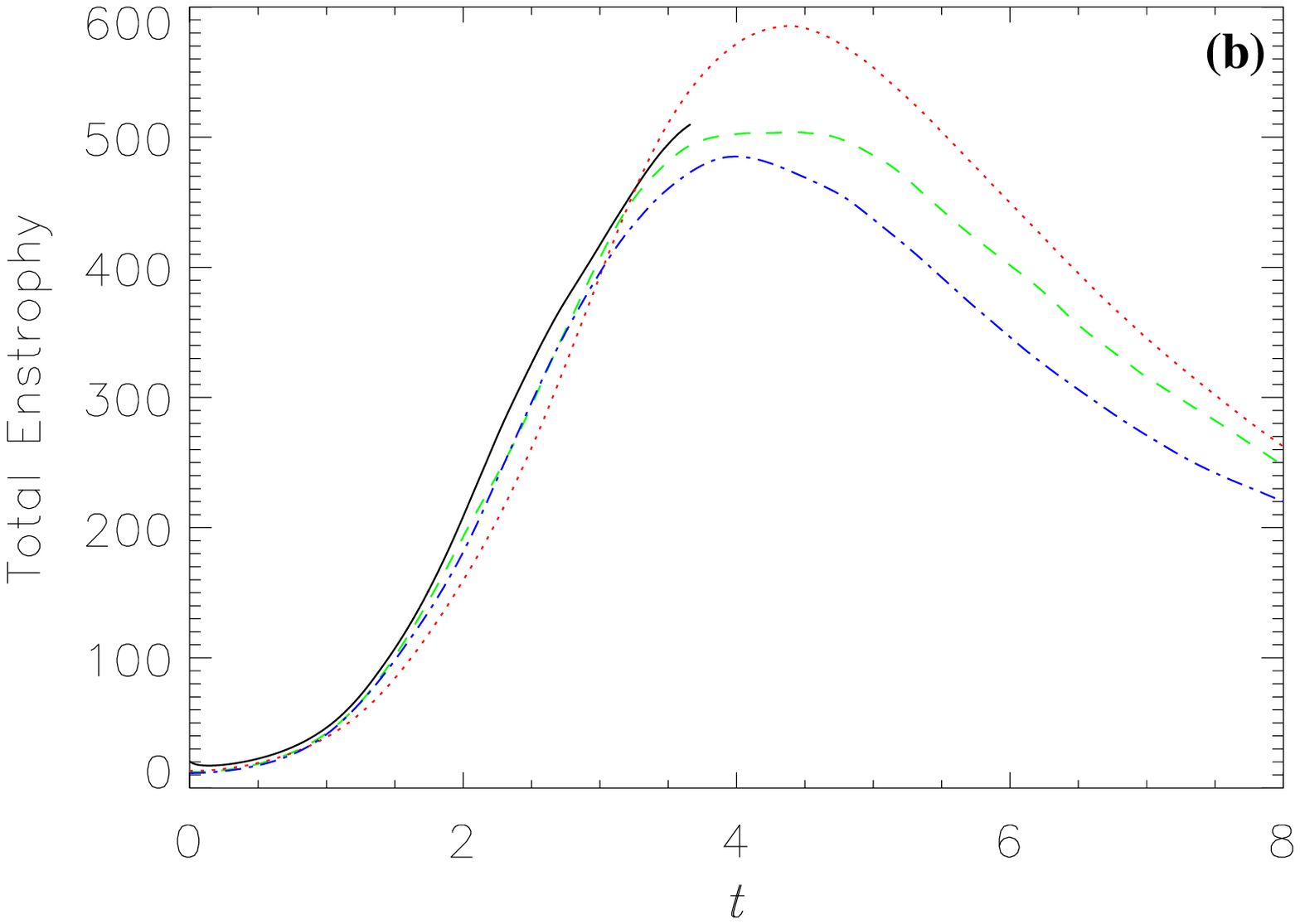}
\end{tabular}}
  \caption{(Color online) Temporal evolution, $\tau_{eddy} \approx 4.5$, for $1536^3$ DNS (solid,
    black), $256^3$ $k_\alpha = 33$ \lamhda (dashed, green online),
     $256^3$ under-resolved ``DNS'' (dotted, red online),
    \neu{and $384^3$ $k_\alpha = 33$ nonhyperdiffusive-\lamhda (dash-dotted, blue online).}
      {\bf (a)} Time evolution of the energies: kinetic (lower curves),
    magnetic (middle curves) and total (upper curves).  {\bf (b)} Time
    evolution of total enstrophy, $\left<j^2+\omega^2\right>$
    ($\left<j^2+\boldsymbol{\omega}\cdot\bar{\boldsymbol{\omega}}\right>$ for \lamhda \neu{and $\left<\vec{j}\cdot\bar{\vec{j}}+\boldsymbol{\omega}\cdot\bar{\boldsymbol{\omega}}\right>$ for the nonhyperdiffusive case).}  Note that \lamhda
    gives a better agreement to the total dissipation rate up to the maximum time that the high resolution DNS is performed.
   Also note that the DNS equivalent to the \lamhda run presented here is not feasible on present-day computers at a reasonable cost.}
     \label{FIG:EVST} \end{center} \end{figure}

\begin{figure}[htbp]\begin{center}\leavevmode
\centerline{%
  \begin{tabular}{c@{\hspace{.15in}}c}
  \includegraphics[width=8.95cm]{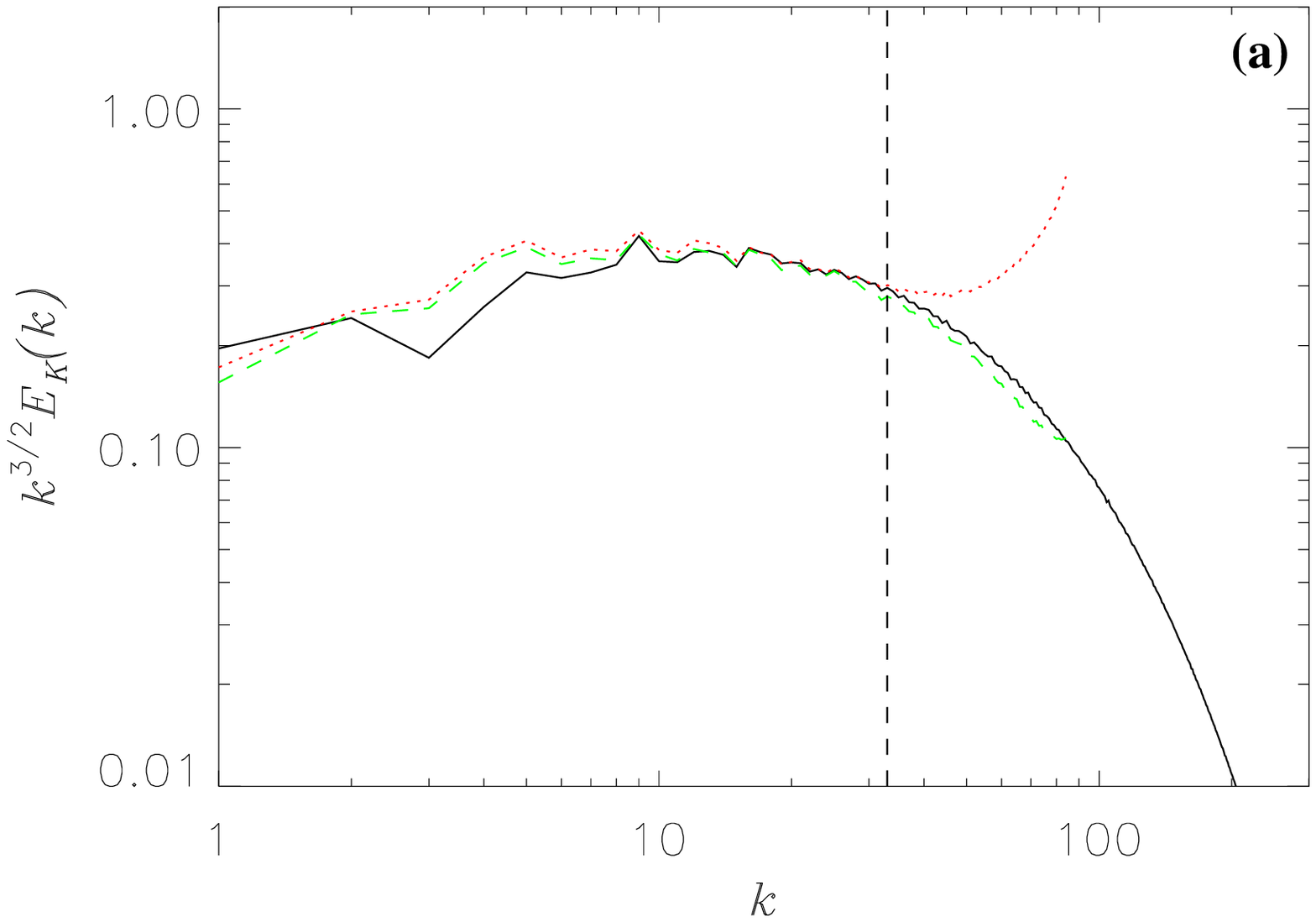} &
  \includegraphics[width=8.95cm]{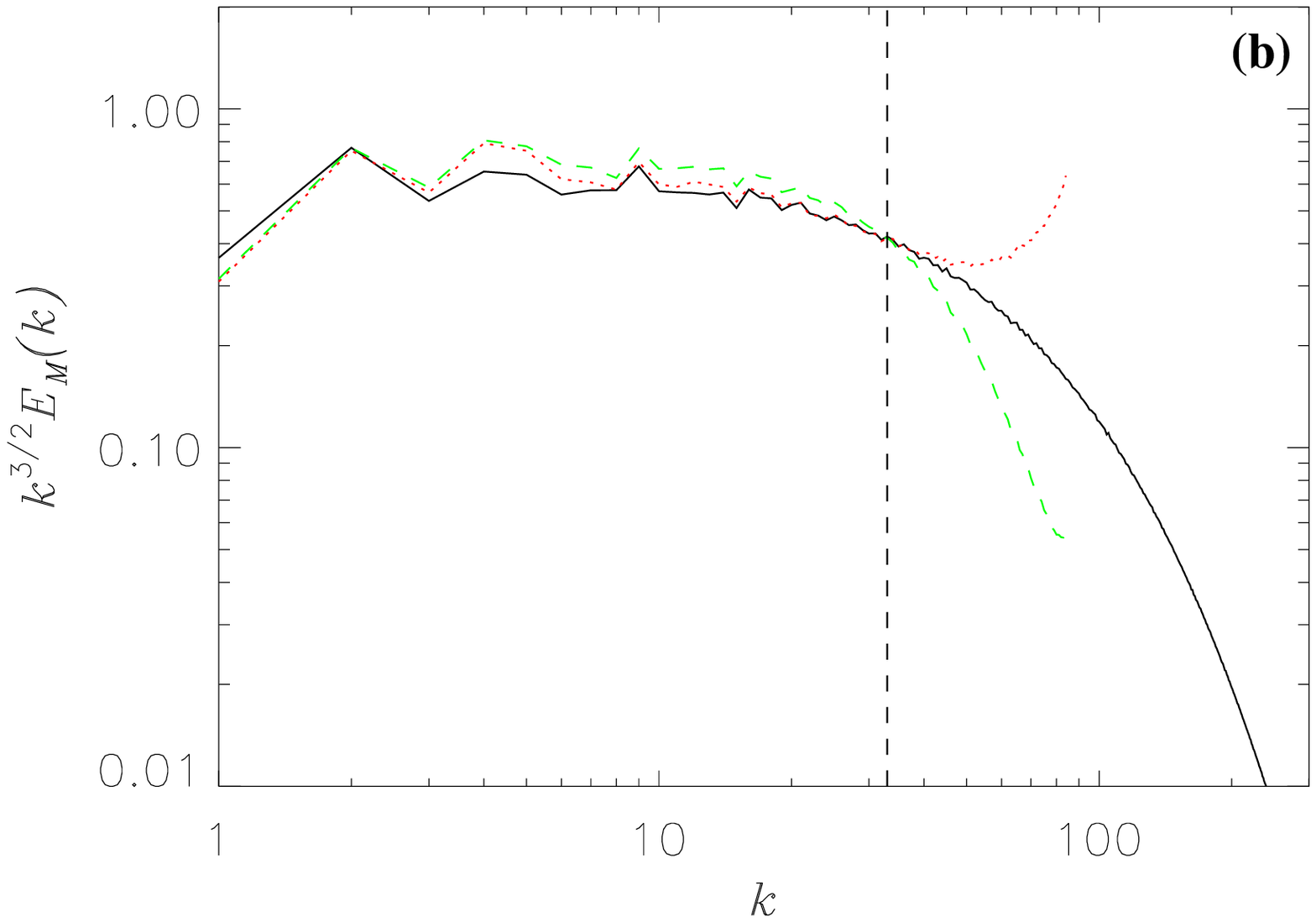}
\end{tabular}}
  \caption{
  Spectra compensated by $k^{3/2}$ for the kinetic {\bf (a)} and magnetic {\bf (b)} energies \neu{averaged over $t \in [2.7,3.7]$;} labels are as in Fig. \ref{FIG:EVST} and the dashed vertical line indicates
    $k_\alpha=33$.  \neu{Note}
    the $k^2$ tail at high wavenumber that is known to develop for under-resolved runs, a prediction stemming from statistical mechanics.
    }
  \label{FIG:COMP2} \end{center} \end{figure}

Having now shown that \lamhda does not suffer the same drawbacks with
regards to energy spectra as \lans, we may turn our attention to \resp{}{a
practical application.}  The purpose of a SGS model or LES is to make predictions
about large Reynolds number flows at a reduced computational expense.
From the scaling arguments in Refs. \cite{FHT01,PGHM+07a}, using
simulations conducted at $Re \approx 2200$, and assuming a
$k^{-1}$ scaling, we can estimate $\alpha=1/33$ for a $256^3$
\lamhd-LES ``prediction'' of our $1536^3$ MHD-DNS.  Time evolution of
the energies and the total enstrophy are shown in Fig. \ref{FIG:EVST}
for much later times than reasonably attainable with the MHD DNS with present-day computers.  Also shown are
results for solving the MHD equations, Eqs. (\ref{eq:mhd}) 
with $\nu =2 \cdot 10^{-4}$ and a resolution of $256^3$: a so-called ``unresolved
DNS'' \resp{}{and the non-hyperdiffusive modified-\lamhda from the previous section.}  Before the peak of dissipation, $t\approx4$, the unresolved
DNS gives a poorer prediction of the total dissipation and total
energy which is then followed by a 
significantly larger and 
somewhat later peak of
dissipation, at $t\approx5$ than the resolved DNS and the \lamhda LES.
\resp{}{The non-hyperdiffusive \lamhda is not expected to perform well as a
SGS model and it is seen to be clearly under-dissipative.  The ratio of magnetic
to kinetic dissipation is $\approx1.5$ for the DNS, $\approx2.9$ for \lamhd, $\approx1.1$ for the under-resolved DNS, and $1.4$ for the non-hyperdiffusive model.}  \resp{}{Together with Fig. \ref{FIG:EVST} (b) these ratios show that
\lamhda achieves accurate total dissipation by an excess of magnetic dissipation and a reduction of kinetic dissipation (both at the small scales).  This
feature has already been depicted in Fig. 15 of Ref. \cite{MMP05a}.}
Compensated energy spectra for \resp{}{the peak of dissipation} ($t\in[2.7,3.7]$) are shown in
Fig.  \ref{FIG:COMP2}.  For the under-resolved DNS, \resp{}{we observe} the appearance of a tail at large wavenumbers with a $k^2$ spectrum as predicted using statistical mechanics arguments for truncated systems in the ideal ($\nu=0$, $\eta=0$) case \cite{FrPoLe+1975}.
The under-resolved spectra are not significantly different from the resolved DNS, but note that a reliable and convincing determination of spectral indices, beyond visual inspection, does require high resolutions.
Comparing now the resolved DNS and the \lamhda run, the quality of the spectra are similar for scales
larger than $\alpha$.  Recall that differences at the largest scales, stem from the
 differences in initial conditions as stated in
Section \ref{SEC:LAMHDGOOD}, and from time evolution of the flow.
Finally, noting that the computer saving here is $6^3$ in memory and $6^4$ in running time, we conclude that the \lamhda continues to behave satisfactorily, as already shown both in two space dimensions 
\cite{MMP05a,PGMP05,PGHM+06}
and in 3D \cite{MMP05b}, 
in particular in the context of the dynamo problem of generation of magnetic fields by velocity gradients; thus, \lamhda may prove to be a useful tool in many astrophysical contexts where magnetic fields are dynamically important, such as in the solar and terrestrial environments, or in the interstellar and intergalactic media.


\begin{figure}[htbp]\begin{center}\leavevmode
\centerline{%
  \begin{tabular}{c@{\hspace{.15in}}c}
  \includegraphics[width=8.95cm]{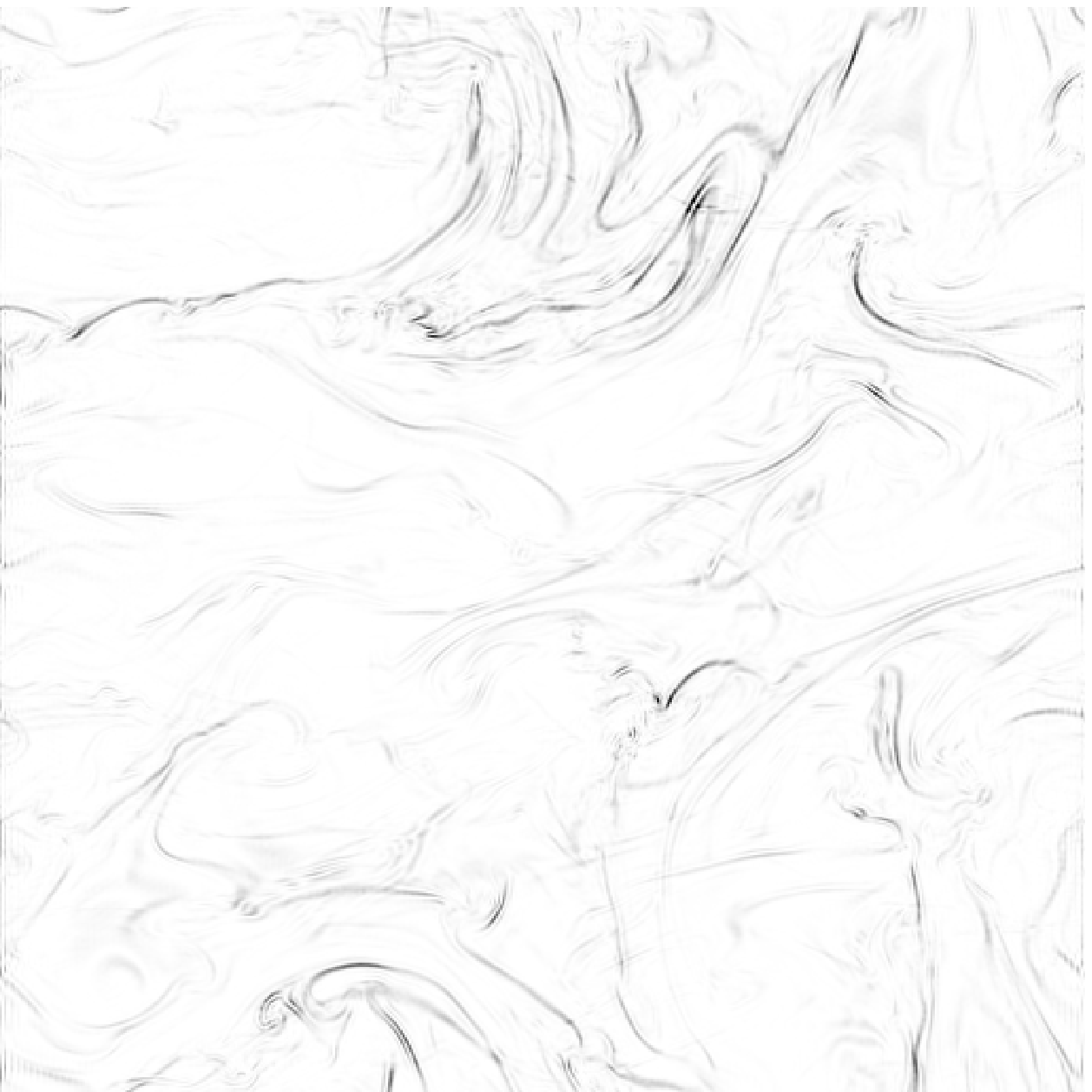} &
  \includegraphics[width=8.95cm]{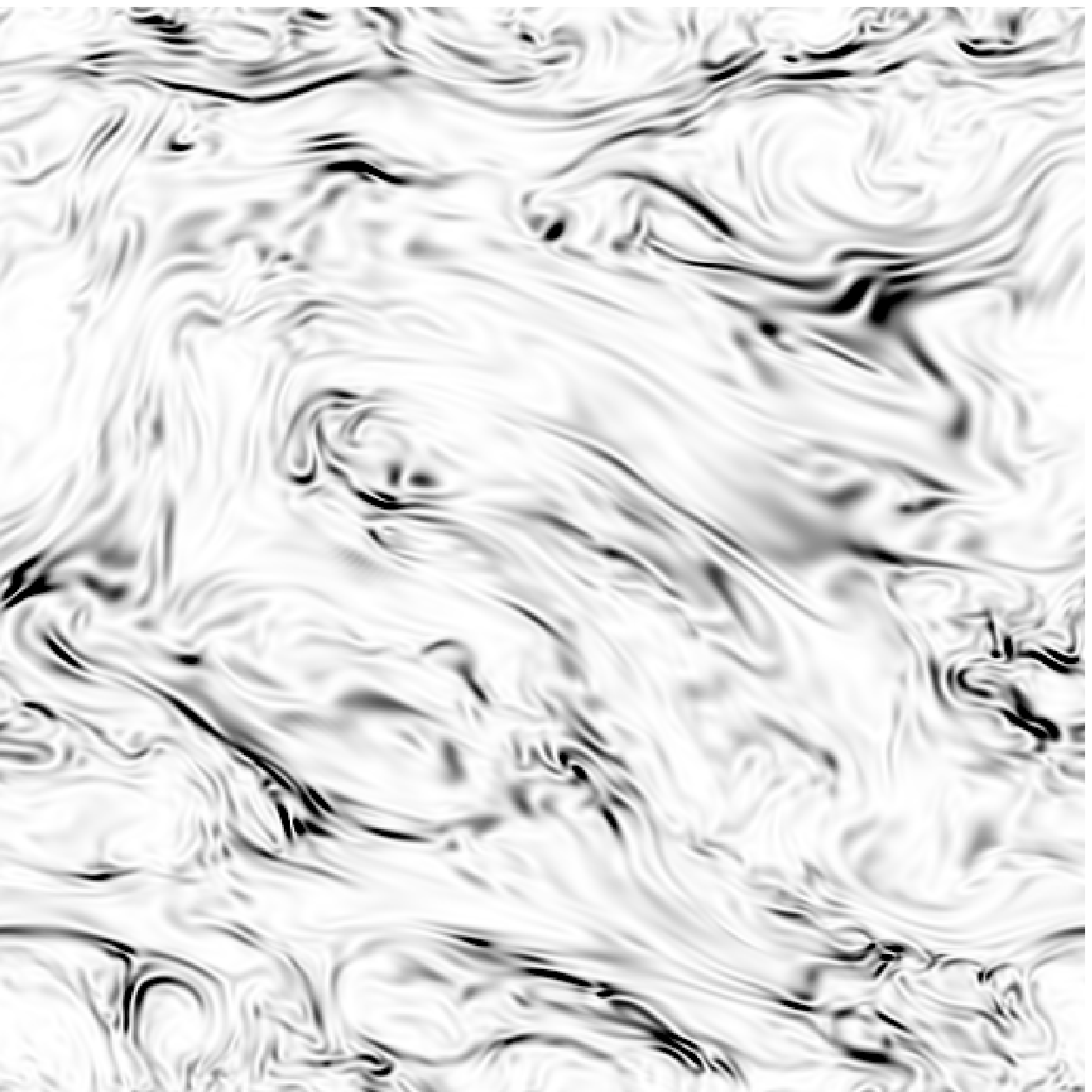} \\
  \includegraphics[width=8.95cm]{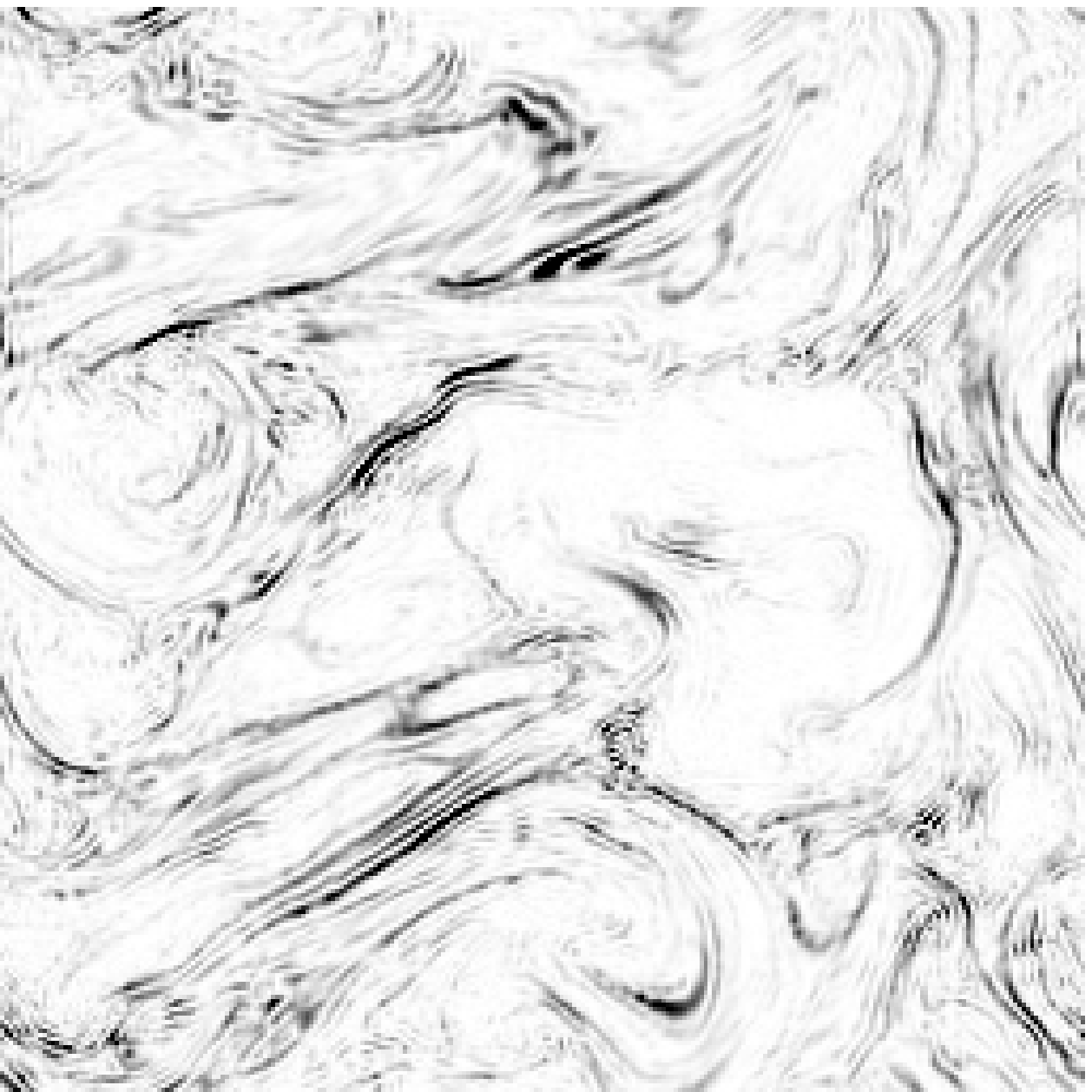} &
  \includegraphics[width=8.95cm]{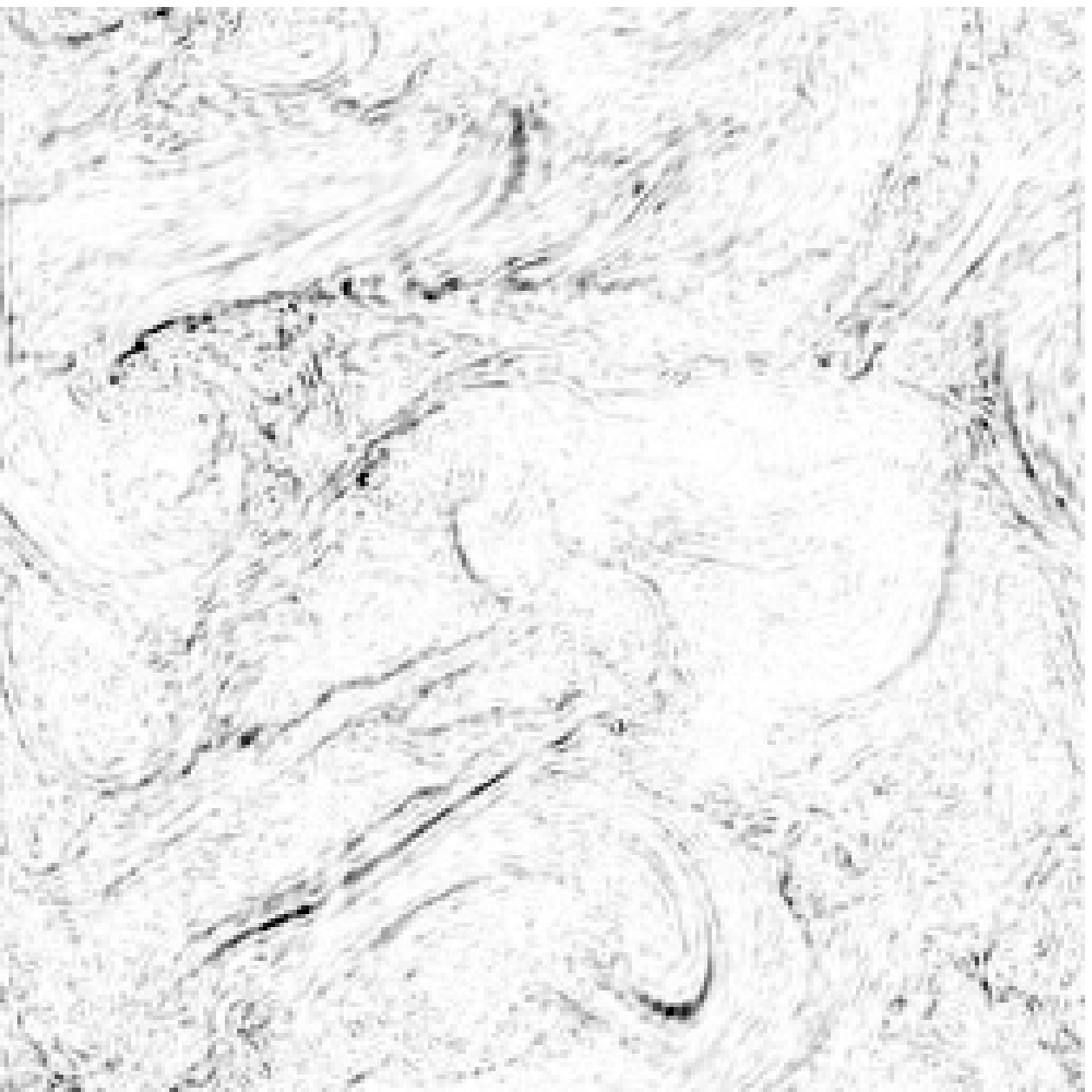} \\
\end{tabular}}
  \caption{\vier{2D cross sections of square current, $j^2$, for
      $512^3$ \lamhd-LES ($\alpha=1/85$) {\bf (Upper Left)} and
      model-stress-case ($\alpha=1/18$) {\bf (Upper Right)}.  MHD
      dissipative structures, sheets, are retained which become
      thicker as $\alpha$ is increased. {\bf (Lower Left)} $256^3$
      \lamhd-LES ($\alpha=1/33$) and {\bf (Lower Right)} $256^3$
      unresolved DNS.  For the unresolved run, current sheets are
      somewhat smeared out by numerical noise.}}
  \label{FIG:SHEETS} \end{center} \end{figure}

\vier{We also computed a $512^3$ \lamhd-LES ($\alpha=1/85$) which retains
more of the small scales than the $256^3$ \lamhd-LES while still
yielding significant computational savings over the $1536^3$ DNS.  We
compare this with the result for $\alpha=1/18$ (chosen not as a LES but
to stress the model) in Fig. \ref{FIG:SHEETS}.  The structure of
sheets observed in MHD dissipative structures is preserved in the
\lamhda simulations, although current and vortex sheets become thicker
in \lamhda as a result of the filter as $\alpha$ is increased.  This
is necessary to achieve reduced resolution computations.  Note that
these sheets are different in nature from the fat 'rigid bodies'
observed in \lans, as the turbulent energy transfer to small scales is
not quenched and there is no super-filter-scale bottleneck.}

\section{Discussion}

In this paper, we have tested the \lamhda model against high Reynolds
number direct numerical simulations (up to \resp{}{Reynolds numbers of
  $\approx 9200$)} and in particular we have focused our attention on
the dynamics of small scales near the $\alpha$ cut-off. We find that
the small-scale spectrum presents no particular defect; specifically,
we find that, unlike in the hydrodynamical case, the
Lagrangian-averaged modeling for MHD exhibits, even at large Reynolds
numbers, neither a positive-power-law spectrum nor any contamination
of the super-filter-scale spectral properties.  \resp{}{This
  difference between \lansa and \lamhda} is not due to the inclusion
of a hyper-diffusive term in \lamhda that stems from the derivation of
the model; rather, it stems from fundamental differences between
hydrodynamics and MHD. Indeed, neither the (non-consistent) removal of
hyperdiffusion from \lamhda nor the examination of scales much smaller
than $\alpha$ gave any indication of problems similar to those caused
by the zero-flux regions found in computations using \lans.  These
regions limited the computational gains of using \lansa as a LES in
hydrodynamics to a factor of only $10$ in computational degrees of
freedom or $30$ in computation time.  \lamhda is not subject to the
same limitations and, as we demonstrated, a gain of a factor of $200$
in the number of degrees of freedom, or a factor of $1300$ in computation time, obtains when comparing to
the highest Reynolds number in turbulent MHD available today in a DNS.

There are two obvious candidates to explain the lack of a
(super-filter-scale) bottleneck effect in \lamhd: the enhanced
(hyper-)diffusion in \lamhda compared with \lans, and \resp{}{physical
  differences between fluids and magneto-fluids, specifically,
  spectrally nonlocal transfer via Alfv\'en waves and its associated} breaking of
the circulation conservation.  The first candidate would eliminate the
super-filter-scale bottleneck by removing energy from the system and
precluding the formation of a secondary range below the filtering
scale $\alpha$ (note that this term becomes of the same order as the
ordinary diffusion when $l\sim\alpha$).  Simulations of \lamhda
performed without the hyper-diffusion term \resp{}{ruled out} this
scenario, as no super-filter bottleneck was found.

The second candidate is the \resp{}{presence of the Lorentz force in
  MHD (and \lamhd) which} breaks down the circulation conservation
\resp{}{and provides the restoring force for Alfv\'en waves.  Both
  properties were shown to be preserved by \lamhd.  In Navier-Stokes,
  the development of helical filaments could quench local interactions
  \cite{MT92,T01} depleting the energy transfer and leading to the
  viscous bottleneck.  However, in MHD, the conservation of the
  circulation ($d\Gamma/dt=0$ in the absence of dissipation) is broken
  by the Lorentz force, which modifies Kelvin's theorem (see
  Eq. (\ref{eq:kelvin})).  The forcing term is associated with the
  Alfv\'en waves, and represents the removal of circulation (and of
  kinetic energy) that is transfered to the magnetic field. Note that
  in Fourier space, the term scales as $k E_M(k)$ and is dominant compared to
  the dissipation in the inertial range. This term precludes the
  formation of rigid bodies, giving as a result a larger net flux
  towards smaller scales and a resulting larger dissipation in
  MHD/\lamhd. This is illustrated in Fig. \ref{FIG:PDFS}. This sink of
  circulation may also be the cause of the lack of a viscous-scale
  bottleneck in MHD. In LANS it was shown
  \cite{PGHM+07a,PiGrHoMi+2008} that conservation of the circulation
  (except for viscosity) leads to the formation of rigid bodies that
  fill a substantial volume of the fluid, and that in turn
  substantially decrease the energy flux to small scales, reduce
  dissipation, and create the super-filter scale bottleneck.} In
\lamhd, the destruction of sub-filter-scale rigid bodies by large
scale magnetic field and shear \resp{}{results} as the presence of a
magnetic field permits the development of long-range interactions in
spectral space \cite{MAP05,AMP05a,Al2007a}.  This can also explain why
$\alpha-$models for other non-local equations, or for problems that do
not preserve the circulation provide good SGS models.  As an example,
the use of \lansa in primitive equations ocean modeling gives
satisfactory results, e.g. in its reproducing the Antarctic
circumpolar current baroclinic instability that can be seen only at
substantially higher resolutions when using direct numerical
simulations \cite{HeHoPe+2008}. 

\resp{}{Energy is dissipated in MHD flows through two different
  processes. Viscosity is responsible for the dissipation of
  mechanical energy, while Ohmic losses are responsible for
  dissipation of magnetic energy. Mechanical and magnetic energy are
  not conserved separately, but rather coupled as illustrated by the
  existence of Alfv\'en waves, which correspond to oscillations of the
  magnetofluid with the velocity field parallel or anti-parallel to
  the magnetic field, and associated to the interchange of magnetic
  and kinetic energy.  In MHD, it is believed that most of the total
  energy in the flow is finally dissipated (mediated by this
  interchange) through Ohmic losses, in a process that involves
  reconnection of magnetic field lines. This is supported by several
  simulations of MHD turbulence \cite{HaBrDo2003,Mi2007a} and is consistent
  with phenomenology. While in hydrodynamics small scales are
  permeated by a myriad of vortex filaments, in MHD the dominant
  dissipative structures are current sheets, where strong gradients of
  the magnetic field and their associated strong currents lead to
  rapid Ohmic dissipation.  Sub-grid models attempt to replace the
  physical processes of small-scale dissipation by processes that
  mimic the non-linear transfer of energy to smaller scales (where
  energy is in reality dissipated, but now in scales that are not
  resolved by the model). In traditional LES, this is done with
  enhanced turbulent viscosities. Note that the eddy viscosity is not
  obtained from the linear dissipative term (the term that describes
  the actual physical process responsible for the dissipation) but
  from the non-linear terms in the equations (the terms that describe
  the coupling between fields at different scales). The final goal is
  not to capture the dissipation processes, but to be able to preserve
  (with computational gains) the large scale dynamics.}

\resp{}{Lagrangian averaged models take a different (although related,
  see e.g., \cite{PGHM+06}) approach. Besides adding (in some cases,
  as in the case of MHD) an enhanced viscosity, the non-linear terms
  are modified at small scales. This modification changes the
  time-scale of the energy cascade, and as a result changes the
  scaling law of the energy spectrum $E(k)$ at sub-filter scales. This
  change leads to changes in the dissipation, as the dissipation is in
  the original equations proportional to $k^2E(k)$. The end result (an
  enhanced dissipation that is intended to mimic the transfer of
  energy to smaller scales in the unresolved scales) should be the
  same as in a traditional LES: gains in computing costs preserving as
  much information of the large scale flow as possible.  As in the
  case of LES, the actual dissipation process is not as important as
  the fact that large-scale dynamics should be reproduced with minimal
  contamination \vier{by} the sub-grid model. We believe the results
  presented here (and in earlier work
  \cite{MMP05a,MMP05b,PGMP05,PGHM+06,Mi2006a}) show this is the case,
  and allow the use of the LAMHD equations as a subgrid model of MHD
  turbulence. However, considering the differences observed between
  LANS and LAMHD, we discuss the dissipation processes in LAMHD.  Two
  mechanisms for dissipation can be identified in LAMHD: dissipation
  of mechanical energy through the viscosity, and dissipation of
  magnetic energy through (enhanced) Ohmic losses. From the equations,
  the total variation of energy goes as \cite{MMP05a}: $dE/dt =
  -\nu\left<\boldsymbol{\omega}\cdot\bar{\boldsymbol{\omega}}\right>
  -\eta\left<j^2\right>$ and as a result the mechanical energy
  dissipation scales as $k^2E_V(k)$ while the magnetic energy
  dissipation scales as $(1+ \alpha^2k^2)k^2E_M(k)$. The extra $k^2$
  factor in the latter gives more dissipation than in the LANS
  case. This excess of magnetic dissipation in LAMHD mimics, as
  previously mentioned, the dominant contribution to dissipation by
  Ohmic losses in MHD. This hyperdiffusion is required in the
  sub-filter scales to accurately model the total energy dissipated at
  the unresolved scales.  This was demonstrated by our experiments
  with a modified \lamhd, where we (non-consistently) removed the
  hyperdiffusive term and found the resulting model to fail as a LES.}

\resp{}{Yet} another way to understand the differences between \lansa
(for incompressible isotropic and homogeneous flows) and \lamhda is to
consider the derivation of these models \cite{H02a} using the
generalized Lagrangian-mean (GLM) formalism \cite{AM78}.  This form of
Lagrangian averaging describes wave, mean-flow interactions.  For the
case of weak turbulence, where the nonlinear transfer is dominated by
waves, GLM requires in principle no closure. As a result, GLM gives an
exact closed theory for the evolution of the wave activity. On the
other hand, when there are no waves (as in incompressible
Navier-Stokes) or when eddies dominate the transfer, a closure is
required.  One possible closure assumes that fast fluctuations are
just advected by the mean flow (basically, Taylor's frozen-in
hypothesis for the small scale turbulent fluctuations) and leads to
the several "$\alpha$-models" that include \lansa and \lamhd.  In this
context, it is not surprising for subgrid models based on GLM to
perform better in the presence of Alfv\'en waves (for \lamhd) or
Rossby and gravity waves (for the Lagrangian-averaged primitive
equations \cite{HeHoPe+2008}).  The more relevant the waves are to the
dynamics, and to the non-linear coupling of modes in the system, the
less relevant is the hypothesis behind the closure.  Furthermore, the
$\alpha$-model equations can then be expected to be a better
approximation to the problem at hand, that is, to be closer to an
exact closure of the original system of equations.

\vier{In the fluid case, the application of the ``Taylor'' closure
  that smaller-than-$\alpha$ scale fluctuations are swept along by the
  large-scale flow results in the fluctuations having greatly reduced
  interactions.  This allows for a reduction in computational expense
  and leads to the super-filter-scale bottleneck by quenching
  spectrally non-local interactions.  In the \lamhda case, the
  small-scale $\vec{z}^+$ ($\vec{z}^-$) fluctuations are swept along
  by the large-scale $\bar{\vec{z}}^-$ ($\bar{\vec{z}}^+$) flow.
  Small-scale fluctuations advected by two different fields may now
  collide and nonlinearly interact.  The second part of the model is
  the preferential hyperdiffusion of Alfv\'en waves
  with wavelengths shorter than $\alpha$.  This damps rather than
  quenches nonlinear interactions among the small scales.  This more
  gentle suppression of the transfer of energy to smaller scales
  reduces the numerical resolution requirements without forming a
  bottleneck.}

It was noted in \cite{MMP05b} when assessing the properties of \lamhda in the dynamo context that the overall temporal evolution was satisfactory, e.g. with a correct growth rate, although
 the growth of the magnetic seed field started slightly earlier in the \lamhda run than in the DNS. One can speculate as to whether this delay is linked to the super-bottleneck effect of \lansa (which prevails when the magnetic field is negligible compared to the velocity, the two modeling approaches, \lamhda and \lans, being dynamically consistent).  This point is left for future work; one could determine as well at what ratio of magnetic to
kinetic energy the overshooting of spectra in \lansa disappears for \lamhd.
 
Also deserving of a separate study is to investigate the behavior of \lamhda when anisotropies that appear at small scales \cite{MiPo2007a} are present; this would be essential when a uniform magnetic field is imposed to the overall flow. The evaluation of the behavior of the model when computing spectra in the perpendicular and parallel directions (with respect to a quasi-uniform magnetic field, computed by locally averaging the field in a sphere of radius comparable to the integral scale) remains to be done but is somewhat time consuming. 
An analysis of the structures that develop in the highly turbulent \lamhda flow studied in the preceding section is also left for future work; of particular interest is the occurrence of Kelvin-Helmholtz like roll-up of current sheets as observed at high resolution \cite{MiPo2007a}; however, the choice of the parameter $\alpha$ in the present paper was made on the basis of questioning the existence or lack thereof of a rigid-body high-wavenumber $k^{+1}$ spectrum and, thus, was not optimized for the study of the inertial range properties of the flow for which a much smaller value of the length $\alpha$ could be used.

Finally, how far resolution can be reduced when using \lamhda as a LES for
various statistics of interest will also require further detailed study. The present study shows that, to reproduce the super-filter-scale energy spectrum in three dimensions, gains by a factor of 1300 in computing time can be achieved. The need to reproduce higher order statistics can decrease these gains. As an example, in two-dimensional MHD, it was shown that gains when using \lamhda as a subgrid model depend for high order moments on the order that 
one wants to see
to be accurately reproduced \cite{PGHM+06}.


\begin{acknowledgments}
Computer time was provided by GWDG, NCAR, and the National Science
Foundation Terascale Computing System at the Pittsburgh Supercomputing
Center.  The NSF Grant No. CMG-0327888 at NCAR supported this work in
part and is gratefully acknowledged.  \add{}{PDM is a member of the
  Carrera del Investigador Cient\'{\i}fico of CONICET.} \resp{}{The
  anonymous referees are gratefully acknowledged for improving the
  clarity of the discussion of our results.}
\end{acknowledgments}


\end{document}